\documentclass[smallextended]{svjour3}

\newif\ifpaper\papertrue

\usepackage{manyfoot}
\usepackage[most]{tcolorbox}
\usepackage[title]{appendix}
\usepackage{balance}
\usepackage{algorithm}
\usepackage{algorithmicx}
\usepackage[caption=false,font=normalsize,labelfont=sf,textfont=sf]{subfig}
\usepackage{stfloats}
\usepackage{capt-of}
\usepackage{graphicx}
\usepackage{xcolor}
\usepackage{tikz}
\usetikzlibrary{shapes.geometric, arrows, fit, shadows}
\usepackage[framemethod=tikz]{mdframed}

\usepackage[paperwidth=178mm,paperheight=254mm,textwidth=119mm,textheight=194mm,hcentering,vcentering]{geometry}

\usepackage[english]{babel}
\usepackage[utf8]{inputenc}
\usepackage[T1]{fontenc}

\usepackage[hyphens]{url}
\usepackage[square,numbers,sort&compress]{natbib}
\usepackage{hyperref}
\usepackage{csquotes}

\usepackage{xspace}
\usepackage[super]{nth}
\usepackage[inline]{enumitem}

\usepackage{amsmath}
\usepackage{amssymb}

\usepackage{array}
\usepackage{tabularx}
\usepackage{booktabs}

\usepackage{listings}
\usepackage{textcomp} 

\lstnewenvironment{plain}{\lstset{numbers=none}}{}

\newcommand*{\enq}[1]{\enquote{{\itshape#1}}}

\newcommand{\must}{\textbf{must}\xspace}
\newcommand{\mustnot}{\textbf{must not}\xspace}
\newcommand{\should}{\textbf{should}\xspace}
\newcommand{\shouldnot}{\textbf{should not}\xspace}

\ifpaper
  \newcommand{\summary}{\underline{\emph{Summary}}\xspace}
\else
  \newcommand{\summary}{\textbf{\emph{Summary}}\xspace}
\fi

\ifpaper
  \newcommand{\paper}{\emph{paper}\xspace}
  \newcommand{\supplementarymaterial}{\emph{supplementary material}\xspace}
\else
  \newcommand{\paper}{\emph{paper}\xspace}
  \newcommand{\supplementarymaterial}{\emph{supplementary material}\xspace}
\fi

\ifpaper
  \newsavebox{\iconMbox}
  \savebox{\iconMbox}{\tikz[baseline=-0.5ex]\draw[fill=black] (0,0) circle (0.3em);}
  \newcommand{\iconM}{\usebox{\iconMbox}\xspace}
  \newsavebox{\iconSbox}
  \savebox{\iconSbox}{\tikz[baseline=-0.5ex]\draw[fill=gray!40, draw=gray!60] (0,0) circle (0.3em);}
  \newcommand{\iconS}{\usebox{\iconSbox}\xspace}
\else
  \newcommand{\iconM}{●\xspace}
  \newcommand{\iconS}{○\xspace}
\fi

\ifpaper
  \newcommand{\condition}[1]{\texttt{\small[#1]}\,}
\else
  \newcommand{\condition}[1]{@@cond:#1@@}
\fi

\ifpaper
  \definecolor{framebg}{gray}{0.95}
  \newmdenv[
    leftmargin=0pt,
    rightmargin=0pt,
    usetwoside=false,
    innerleftmargin=6pt,
    innerrightmargin=6pt,
    innertopmargin=4pt,
    innerbottommargin=4pt,
    skipabove=0.5\baselineskip plus 1pt,
    skipbelow=0.5\baselineskip plus 2pt,
    linecolor=black,
    backgroundcolor=framebg,
    linewidth=2pt,
    roundcorner=0pt,
    nobreak=false,
    shadow=false,
    topline=false,
    bottomline=false,
    rightline=false,
    leftline=true,
  ]{framed}
\else
  \newenvironment{framed}{\begin{quote}}{\end{quote}}
\fi

\ifpaper
  \newcommand{\scope}{\hyperref[sec:scope]{\mbox{\emph{Scope}}}\xspace}
\else
  \newcommand{\scope}{\href{/scope/}{\emph{Scope}}\xspace}
\fi

\ifpaper
  \newcommand{\studytypes}{\hyperref[sec:study-types]{\mbox{\emph{Study Types}}}\xspace}
  \newcommand{\llmsforresearcher}{\hyperref[sec:llms-as-tools-for-software-engineering-researchers]{\mbox{\emph{LLMs for Research}}}\xspace}
  \newcommand{\annotators}{\hyperref[sec:llms-as-annotators]{\mbox{\emph{LLMs as Annotators}}}\xspace}
  \newcommand{\judges}{\hyperref[sec:llms-as-judges]{\mbox{\emph{LLMs as Judges}}}\xspace}
  \newcommand{\synthesis}{\hyperref[sec:llms-for-synthesis]{\mbox{\emph{LLMs for Synthesis}}}\xspace}
  \newcommand{\subjects}{\hyperref[sec:llms-as-subjects]{\mbox{\emph{LLMs as Subjects}}}\xspace}
  \newcommand{\llmsforengineers}{\hyperref[sec:llms-as-tools-for-software-engineers]{\mbox{\emph{LLMs for SE}}}\xspace}
  \newcommand{\llmusage}{\hyperref[sec:studying-llm-usage-in-software-engineering]{\mbox{\emph{Studying LLM Usage}}}\xspace}
  \newcommand{\newtools}{\hyperref[sec:llms-for-new-software-engineering-tools]{\mbox{\emph{LLMs for Tools}}}\xspace}
  \newcommand{\benchmarkingtasks}{\hyperref[sec:benchmarking-llms-for-software-engineering-tasks]{\mbox{\emph{Benchmarking LLMs}}}\xspace}
\else
  \newcommand{\studytypes}{\href{/study-types/}{\emph{Study Types}}\xspace}
  \newcommand{\llmsforresearcher}{\href{/study-types/llms-for-research/}{\emph{LLMs for Research}}\xspace}
  \newcommand{\annotators}{\href{/study-types/annotators/}{\emph{LLMs as Annotators}}\xspace}
  \newcommand{\judges}{\href{/study-types/judges/}{\emph{LLMs as Judges}}\xspace}
  \newcommand{\synthesis}{\href{/study-types/synthesis/}{\emph{LLMs for Synthesis}}\xspace}
  \newcommand{\subjects}{\href{/study-types/subjects/}{\emph{LLMs as Subjects}}\xspace}
  \newcommand{\llmsforengineers}{\href{/study-types/llms-for-se/}{\emph{LLMs for SE}}\xspace}
  \newcommand{\llmusage}{\href{/study-types/usage/}{\emph{Studying LLM Usage}}\xspace}
  \newcommand{\newtools}{\href{/study-types/tools/}{\emph{LLMs for Tools}}\xspace}
  \newcommand{\benchmarkingtasks}{\href{/study-types/benchmarking/}{\emph{Benchmarking LLMs}}\xspace}
\fi

\ifpaper
  \newcommand{\guidelines}{\hyperref[sec:guidelines]{\mbox{\emph{Guidelines}}}\xspace}
  
  \newcommand{\modelversion}{\hyperref[sec:report-model-version-configuration-and-customizations]{\mbox{\emph{Version and Configuration}}}\xspace}
  \newcommand{\design}{\hyperref[sec:report-system-and-prompt-design]{\mbox{\emph{System and Prompt Design}}}\xspace}
  \newcommand{\traces}{\hyperref[sec:report-session-traces]{\mbox{\emph{Session Traces}}}\xspace}
  \newcommand{\benchmarksmetrics}{\hyperref[sec:use-suitable-baselines-benchmarks-and-metrics]{\mbox{\emph{Benchmarks and Metrics}}}\xspace}
  \newcommand{\openllm}{\hyperref[sec:use-an-open-llm-as-a-baseline]{\mbox{\emph{Open LLM}}}\xspace}
  \newcommand{\humanvalidation}{\hyperref[sec:use-human-validation-for-llm-outputs]{\mbox{\emph{Human Validation}}}\xspace}
  \newcommand{\limitationsmitigations}{\hyperref[sec:report-limitations-and-mitigations]{\emph{Limitations and Mitigations}}\xspace}
  \newcommand{\refdeclareusage}{\hyperref[sec:declare-llm-usage-and-role]{Declare Usage}}
  \newcommand{\refmodelversion}{\hyperref[sec:report-model-version-configuration-and-customizations]{Model Version}}
  \newcommand{\refdesign}{\hyperref[sec:report-system-and-prompt-design]{Design}}
  \newcommand{\reftraces}{\hyperref[sec:report-session-traces]{Traces}}
  \newcommand{\refbenchmarks}{\hyperref[sec:use-suitable-baselines-benchmarks-and-metrics]{Benchmarks \& Metrics}}
  \newcommand{\refopenllm}{\hyperref[sec:use-an-open-llm-as-a-baseline]{Open LLM}}
  \newcommand{\refhumanvalidation}{\hyperref[sec:use-human-validation-for-llm-outputs]{Human Validation}}
  \newcommand{\reflimitations}{\hyperref[sec:report-limitations-and-mitigations]{Limitations}}
\else
  \newcommand{\guidelines}{\href{/guidelines/}{\emph{Guidelines}}\xspace}
  
  \newcommand{\modelversion}{\href{/guidelines/model-version/}{\emph{Version and Configuration}}\xspace}
  \newcommand{\design}{\href{/guidelines/design/}{\emph{System and Prompt Design}}\xspace}
  \newcommand{\traces}{\href{/guidelines/traces/}{\emph{Session Traces}}\xspace}
  \newcommand{\benchmarksmetrics}{\href{/guidelines/benchmarks-metrics/}{\emph{Benchmarks and Metrics}}\xspace}
  \newcommand{\openllm}{\href{/guidelines/open-llm/}{\emph{Open LLM}}\xspace}
  \newcommand{\humanvalidation}{\href{/guidelines/human-validation/}{\emph{Human Validation}}\xspace}
  \newcommand{\limitationsmitigations}{\href{/guidelines/limitations/}{\emph{Limitations and Mitigations}}\xspace}
  \newcommand{\refdeclareusage}{\href{/guidelines/declare-usage/}{Declare Usage}}
  \newcommand{\refmodelversion}{\href{/guidelines/model-version/}{Model Version}}
  \newcommand{\refdesign}{\href{/guidelines/design/}{Design}}
  \newcommand{\reftraces}{\href{/guidelines/traces/}{Traces}}
  \newcommand{\refbenchmarks}{\href{/guidelines/benchmarks-metrics/}{Benchmarks \& Metrics}}
  \newcommand{\refopenllm}{\href{/guidelines/open-llm/}{Open LLM}}
  \newcommand{\refhumanvalidation}{\href{/guidelines/human-validation/}{Human Validation}}
  \newcommand{\reflimitations}{\href{/guidelines/limitations/}{Limitations}}
\fi

\ifpaper
  \newcommand{\guidelinesubsubsection}[1]{\subsubsection{\bfseries #1:}}
  \newcommand{\studytypesubsection}[1]{\subsubsection{#1}}
  \newcommand{\studytypeparagraph}[1]{\paragraph{\bfseries #1:}}
  \newcommand{\scopeparagraph}[1]{\paragraph{\bfseries #1:}}
  \newcommand{\guidelineparagraph}[1]{\paragraph{#1.}}
\else
  \newcommand{\guidelinesubsubsection}[1]{\subsubsection{#1}}
  \newcommand{\studytypesubsection}[1]{\subsection{#1}}
  \newcommand{\studytypeparagraph}[1]{\subsubsection{#1}}
  \newcommand{\scopeparagraph}[1]{\subsubsection{#1}}
  \newcommand{\guidelineparagraph}[1]{\paragraph{#1}}
\fi

\ifpaper

  \newcommand{\seemodelversion}{\seesection{sec:report-model-version-configuration-and-customizations}}
  \newcommand{\seedesign}{\seesection{sec:report-system-and-prompt-design}}
  \newcommand{\seetraces}{\seesection{sec:report-session-traces}}
  \newcommand{\seebenchmarksmetrics}{\seesection{sec:use-suitable-baselines-benchmarks-and-metrics}}
  \newcommand{\seeopenllm}{\seesection{sec:use-an-open-llm-as-a-baseline}}
  \newcommand{\seehumanvalidation}{\seesection{sec:use-human-validation-for-llm-outputs}}
  \newcommand{\seelimitationsmitigations}{\seesection{sec:report-limitations-and-mitigations}}
\else
  
  \newcommand{\seemodelversion}{\href{/guidelines/model-version/}{Report Model Version, Configuration, and Customizations}}
  \newcommand{\seedesign}{\href{/guidelines/design/}{Report System and Prompt Design}}
  \newcommand{\seetraces}{\href{/guidelines/traces/}{Report Session Traces}}
  \newcommand{\seebenchmarksmetrics}{\href{/guidelines/benchmarks-metrics/}{Use Suitable Baselines, Benchmarks, and Metrics}}
  \newcommand{\seeopenllm}{\href{/guidelines/open-llm/}{Use an Open LLM as a Baseline}}
  \newcommand{\seehumanvalidation}{\href{/guidelines/human-validation/}{Use Human Validation for LLM Outputs}}
  \newcommand{\seelimitationsmitigations}{\href{/guidelines/limitations/}{Report Limitations and Mitigations}}
\fi
\space%

\usepackage{needspace}
\usepackage{etoolbox}
\pretocmd{\section}{\needspace{6\baselineskip}}{}{}
\pretocmd{\subsection}{\needspace{10\baselineskip}}{}{}
\pretocmd{\subsubsection}{\needspace{8\baselineskip}}{}{}

\makeatletter
\def\makeheadbox{%
  \hbox to0pt{%
    \vbox{\baselineskip=10dd
      \hrule
      \hbox to\hsize{%
        \vrule\kern3pt
        \vbox{\kern3pt
          \hbox{\bfseries Empirical Software Engineering}%
          \hbox{(accepted manuscript)}%
          \kern3pt}%
        \hfil\kern3pt\vrule}%
      \hrule}%
    \hss}}
\makeatother

\date{Submitted: 26/08/2025\\Revised: 27/02/2026, 05/05/2026, 11/06/2026}

\title{Guidelines for Empirical Studies in Software~Engineering involving Large~Language~Models}

\author{Sebastian~Baltes \and Florian~Angermeir \and Chetan~Arora \and Marvin~Muñoz~Barón \and Chunyang~Chen \and Lukas~Böhme \and Fabio~Calefato \and Neil~Ernst \and Davide~Falessi \and Brian~Fitzgerald \and Davide~Fucci \and Junda~He \and Christoph~Treude \and Marcos~Kalinowski \and Stefano~Lambiase \and Daniel~Russo \and Mircea~Lungu \and Cristina~Martinez~Montes \and Lutz~Prechelt \and Paul~Ralph  \and Rijnard~van~Tonder  \and Stefan~Wagner}

\institute{%
  Sebastian~Baltes \at
  Heidelberg University, Germany\\
  \email{sebastian.baltes@uni-heidelberg.de}\\
  (corresponding author)
  \and
  Florian~Angermeir \at
  fortiss, Germany\\
  Blekinge Institute of Technology, Sweden\\
  \email{angermeir@fortiss.org}
  \and
  Chetan~Arora \at
  Monash University, Australia\\
  \email{chetan.arora@monash.edu}
  \and
  Marvin~Muñoz~Barón  \and Chunyang~Chen \at
  Technical University of Munich, Germany\\
  \email{\{marvin.munoz-baron, chun-yang.chen\}@tum.de}
  \and
  Lukas~Böhme \at
  Hasso-Plattner-Institut, Germany\\
  University of Potsdam, Germany\\
  \email{lukas.boehme@hpi.de}
  \and
  Fabio~Calefato \at
  University of Bari, Italy\\
  \email{fabio.calefato@uniba.it}
  \and
  Neil~Ernst \at
  University of Victoria, Canada\\
  \email{nernst@uvic.ca}
  \and
  Davide~Falessi \at
  University of Rome Tor Vergata, Italy\\
  \email{falessi@ing.uniroma2.it}
  \and
  Brian~Fitzgerald \at
  Lero, Ireland\\
  University of Limerick, Ireland\\
  \email{brian.fitzgerald@ul.ie}
  \and
  Davide~Fucci \at
  Blekinge Institute of Technology, Sweden\\
  \email{davide.fucci@bth.se}
  \and
  Junda~He \and Christoph~Treude \at
  Singapore Management University, Singapore\\
  \email{\{jundahe, ctreude\}@smu.edu.sg}
  \and
  Marcos~Kalinowski \at
  PUC Rio de Janeiro, Brazil\\
  \email{kalinowski@inf.puc-rio.br}
  \and
  Stefano~Lambiase \and Daniel~Russo \at
  Aalborg University in Copenhagen, Denmark\\
  \email{\{stla, daniel.russo\}@cs.aau.dk}
  \and
  Mircea~Lungu \at
  IT University of Copenhagen, Denmark\\
  \email{mlun@itu.dk}
  \and
  Cristina~Martinez~Montes \at
  Chalmers University of Technology, Sweden\\
  University of Gothenburg, Sweden\\
  \email{montesc@chalmers.se}
  \and
  Lutz~Prechelt \at
  Freie Universität Berlin, Germany\\
  \email{prechelt@inf.fu-berlin.de}
  \and
  Paul~Ralph \at
  Dalhousie University, Canada\\
  \email{paulralph@dal.ca}
  \and
  Rijnard~van~Tonder \at
  Independent Researcher, Antigua and Barbuda\\
  \email{rvantonder@gmail.com}
  \and
  Stefan~Wagner \at
  Technical University of Munich, Germany\\
  \email{stefan.wagner@tum.de}
}

\authorrunning{Baltes et al.}
\titlerunning{Guidelines for Empirical Studies in SE involving LLMs}

\begin{document}

\maketitle

\begin{abstract}
Large Language Models (LLMs) are widely used in software engineering (SE)
research and practice, yet their non-determinism, opaque training data, and
rapidly evolving models threaten the reproducibility and replicability of
empirical studies.
We address this challenge through a collaborative effort of 22 researchers,
presenting a taxonomy of seven study types that organizes how LLMs are
used in SE research, together with eight guidelines for designing
and reporting such studies.
Each guideline distinguishes requirements (must) from recommendations
(should) and is contextualized by the study types it applies to.
Our guidelines recommend that researchers:
(1)~declare LLM usage and role;
(2)~report model versions, configurations, and customizations;
(3)~document the system and prompt design beyond the model;
(4)~report session traces, i.e., interaction logs and runtime traces;
(5)~use suitable baselines, benchmarks, and metrics;
(6)~include an open LLM as a baseline;
(7)~validate LLM outputs against human judgment; and
(8)~articulate limitations and mitigations.
We complement the guidelines with an applicability matrix mapping guidelines
to study types and a reporting checklist for authors and reviewers.
We maintain the study types and guidelines online as a living resource for the
community to use and shape
(\href{https://llm-guidelines.org/}{llm-guidelines.org}).
\space%
\end{abstract}

\keywords{software engineering, large language models, empirical research, meta-science, guidelines}

\section{Introduction}
\label{sec:introduction}

Since the release of ChatGPT in November 2022, large language models (LLMs) have been adopted widely across software engineering (SE) research~\cite{DBLP:journals/tosem/HouZLYWLLLGW24}, yet the reproducibility and replicability of empirical studies involving LLMs remains uncertain.
Recent findings indicate low reproducibility in SE studies involving LLMs.
\citeauthor{DBLP:journals/corr/abs-2510-25506} examined 85 LLM-centric ICSE and ASE 2024 papers~\cite{DBLP:journals/corr/abs-2510-25506}.
Of the 18 papers that both used OpenAI models and provided artifacts, only five were complete enough to execute, and none fully reproduced the original results~\cite{DBLP:journals/corr/abs-2510-25506}.
\citeauthor{DBLP:conf/ndss/EvertzRNMNSGPSS26} found at least one LLM-specific pitfall in each of 72 peer-reviewed security and SE papers from 2023--2024, and only 15.7\% of the observed pitfalls were discussed~\cite{DBLP:conf/ndss/EvertzRNMNSGPSS26}.

LLM-based SE studies are hard to reproduce for three reasons.
First, their inherent non-determinism causes variability across runs~\cite{DBLP:conf/naacl/SongWLL25, DBLP:journals/corr/abs-2602-07150}, and slight changes can lead to substantially different results~\cite{DBLP:conf/nips/AgarwalSCCB21, DBLP:journals/corr/abs-2412-12509}.
Second, commercial models evolve beyond version identifiers, so reported performance can change over time~\cite{DBLP:journals/corr/abs-2307-09009}.
Third, even for ``open'' models, training data and fine-tuning details often remain undisclosed~\cite{Gibney2024}.
Moreover, prompt formatting choices alone can shift accuracy by up to 76 percentage points~\cite{DBLP:conf/iclr/Sclar0TS24}, and configured parameters such as temperature affect output variability~\cite{renze2024temperature}.
Hence, not reporting these settings directly affects reproducibility.

Traditional open science practice in SE has focused on releasing source code and datasets as a replication package. The empirical artifact analysis literature in SE focuses more on code repositories and data than on upstream artifacts such as requirements specifications and design documents~\cite{DBLP:journals/jss/LiuHHXZJCM24}. With LLMs, code is often generated from prompts, context files, and other runtime inputs to the model, so reporting must shift left to cover these upstream artifacts in addition to the code and data they produce.

Although the SE research community has developed guidelines for conducting and reporting specific types of empirical studies such as controlled experiments (e.g.,~\cite{DBLP:books/sp/WohlinRHORW24, DBLP:books/sp/08/SSS2008}), their replications (e.g.,~\cite{DBLP:journals/tse/SantosVOJ21}), or empirical studies in general (e.g., the \emph{ACM SIGSOFT Empirical Standards}~\cite{ralph2021empiricalstandardssoftwareengineering}), none of these address the LLM-specific aspects described above.
Previously, a position paper highlighted these issues~\cite{DBLP:conf/wsese/0001BFB25}, but there was no comprehensive community-developed guidance for designing and reporting empirical studies involving LLMs in SE.

Therefore, we present community-developed guidelines for designing and reporting studies involving LLMs in SE research, co-developed by 22 researchers.
After outlining our \scope, we introduce a taxonomy of \studytypes, then present eight \guidelines.
We complement these with an applicability matrix mapping guidelines to study types and a reporting checklist for authors and reviewers.
For each study type and guideline, we identify relevant examples, both within and outside of SE research.
We maintain the study types and guidelines online as a living resource for the community to use and shape (\href{https://llm-guidelines.org/}{llm-guidelines.org}).
\space%

\section{Scope and Conventions}
\label{sec:scope}

\scopeparagraph{Software Engineering as our Target Discipline}

We target SE research because reporting guidelines from other disciplines do not address its specific needs (see \emph{Related Reporting Guidelines} below). In particular, SE research employs a wide variety of empirical methods~\cite{ralph2021empiricalstandardssoftwareengineering, DBLP:books/sp/WohlinRHORW24}. We organize the LLM-involving subset of these methods into a taxonomy of \studytypes\ifpaper~(Section~\ref{sec:study-types})\fi, and each of our \guidelines specifies its applicability per study type.

\scopeparagraph{Focus on Text-Based Use Cases}

While multi-modal foundation models that use or generate images, audio, or video may also support SE research and practice, we focus on textual use cases of LLMs (e.g., in natural language or programming languages). Many of our guidelines, particularly those concerning model reporting, prompt documentation, and reproducibility, are likely applicable to multi-modal settings as well, but we leave a systematic assessment of this applicability to future work.

\scopeparagraph{Focus on Direct Development or Research Support}

While researchers may use LLMs for many peripheral tasks (e.g., proofreading, spell-checking, translation), our guidelines focus on their direct role in empirical research and engineering practice.
For engineers, we focus on the use of LLMs to automate SE tasks, that is, artificial intelligence (AI) for software engineering (AI4SE) (see~\llmsforengineers).
This includes agentic systems that autonomously plan and execute multi-step tasks using LLMs (see~\design).
For researchers, we focus on the use of LLMs to automate empirical research tasks such as data collection, processing, or analysis (see~\llmsforresearcher).
The 2026 ACM Policy on Authorship likewise requires disclosing AI used in the research itself, not AI used only to assist with writing~\cite{ACM2026}.

\scopeparagraph{Researchers as our Target Audience}

Our guidelines are intended to help SE researchers design, plan, conduct,
and report empirical studies involving LLMs, and to support scholarly peer
review of such studies. Each guideline includes an \emph{Advice for
Reviewers} subsection with targeted assessment suggestions.
Our guidelines focus on what to report and how. They complement but do not replace methodological guidance for designing specific types of empirical studies.

\scopeparagraph{Reporting Locations}

We use \paper to refer to the manuscript PDF, including any appendices it contains. We use \supplementarymaterial to refer to any artifact, replication package, dataset, or other resource external to the manuscript PDF (e.g., hosted on Zenodo or Figshare). What belongs in the main body versus an appendix is a presentation choice for authors and venues; what belongs inside the manuscript versus outside it is the reporting-location decision the guidelines make. When a recommendation does not specify a location, either \paper or \supplementarymaterial is acceptable.

\scopeparagraph{Navigating These Guidelines}

Our guidelines are structured to support different reading strategies depending on the reader's goal.
\emph{Researchers planning a new study} may start with the taxonomy of study types (see \studytypes) to identify which types apply to their planned work, then consult \ifpaper Table~\ref{tab:guideline-matrix} \fi to determine which guidelines are requirements (\must) and which are recommendations (\should) for those study types. Each guideline section opens with a brief summary, allowing readers to quickly assess relevance before reading the full text.
\emph{Researchers writing up results} may start with the \ifpaper checklist in Appendix~\ref{sec:checklist}, \fi which organizes actionable items by typical paper sections (Introduction, Research Design and Methods, Results, etc.). \ifpaper Table~\ref{tab:guidelines} provides a quick reference to all eight guidelines, and Table~\ref{tab:rationale-recommendations} maps each guideline's rationale to its recommendations. \fi
\emph{Reviewers} can use the \emph{Advice for Reviewers} subsection at the end of each guideline for targeted guidance on assessing manuscripts. \ifpaper Table~\ref{tab:guideline-matrix} helps reviewers identify which guidelines apply to the study type under review. \fi

\scopeparagraph{Related Reporting Guidelines}

Reporting guidelines have a long tradition in healthcare research, where CONSORT~\cite{Schulz2010} is the canonical standard for randomized clinical trials. Our reporting checklist (\ifpaper Appendix~\ref{sec:checklist} \fi) follows its structural template. LLM-specific reporting guidelines have appeared more recently. Outside SE, these include \citeauthor{Gallifant2025}'s TRIPOD-LLM for healthcare~\cite{Gallifant2025}, \citeauthor{DBLP:conf/chi/NavarroSA26}'s HCI guidelines~\cite{DBLP:conf/chi/NavarroSA26}, and \citeauthor{Kapoor2024REFORMS}'s REFORMS for ML-based science~\cite{Kapoor2024REFORMS}. Within SE, they include \citeauthor{DBLP:conf/icse/SallouDP24}'s vision paper on validity threats~\cite{DBLP:conf/icse/SallouDP24}, our earlier workshop position paper~\cite{DBLP:conf/wsese/0001BFB25}, and \citeauthor{DBLP:journals/corr/abs-2601-01954}'s prompt-reporting guideline for automated SE~\cite{DBLP:journals/corr/abs-2601-01954}, derived from a literature review of ICSE, FSE, and ASE papers and a survey of 105 program committee members. Beyond LLM-specific work, the NeurIPS reproducibility checklist~\cite{DBLP:journals/jmlr/PineauVSLBdFL21} prescribes per-submission disclosure for ML papers, covering items such as algorithm description, experimental setup, and statistical reporting.

\citeauthor{DBLP:journals/corr/abs-2601-01954} frame their recommendations as complementing ours. The items their 105 surveyed program committee members endorsed as essential align with what \modelversion, \design, and \limitationsmitigations already require. \citeauthor{DBLP:conf/chi/NavarroSA26}'s HCI guidelines, by contrast, balance \enq{the practical realities of authors' time, cost, and page limitations} with scientific concerns and HCI norms, while our \guidelines aim for comprehensive coverage and use \must and \should to express that balance. On prompt reporting, \citeauthor{DBLP:conf/chi/NavarroSA26} argue for selective disclosure based on each prompt's centrality to author claims, whereas \design recommends fuller disclosure with named exceptions for privacy, anonymity, or confidentiality concerns. On technical evaluation, \citeauthor{DBLP:conf/chi/NavarroSA26} recommend a \enq{modest technical evaluation of the LLM component on a dataset of representative inputs} tailored to HCI research, whereas \benchmarksmetrics asks for established benchmarks, traditional baselines, and inferential statistics where applicable. Peripheral LLM uses such as proofreading, spell-checking, and translation are entirely out of scope (see \emph{Focus on Direct Development or Research Support} above).

Our guidelines name \paper or \supplementarymaterial as the reporting location for each recommendation (see \emph{Reporting Locations} above). In summary, our guidelines apply to SE empirical research broadly, organize recommendations around the taxonomy of seven study types introduced in \studytypes with explicit per-study-type applicability, and pair each recommendation with a target reporting location in \paper or \supplementarymaterial.

\section{Methodology}
\label{sec:methodology}

The development of these guidelines was initiated at the 2024 meeting of the \emph{International Software Engineering Research Network} (ISERN) in Barcelona, Spain, where the first and last author organized a session on guidelines for empirical studies involving LLMs. The resulting discussions led to a position paper~\citep{DBLP:conf/wsese/0001BFB25}. A preprint of this paper was the basis for further discussions at the \emph{\nth{2} Copenhagen Symposium on Human-Centered Software Engineering AI} (CHASEAI 2024), where a workstream formed to collaboratively refine and extend the preliminary guidelines.
Our process followed an established tradition of developing reporting guidelines through expert collaboration, as exemplified by CONSORT for clinical trials~\citep{Begg1996} and, more recently, TRIPOD-LLM for LLM-based prediction studies in medicine~\citep{Gallifant2025}.

We held bi-weekly meetings, in which we decided on the structure of the study type taxonomy and the guidelines.
During these discussions, we added a new study type (\synthesis), refined and extended the guideline sections (in particular \modelversion, \design, \traces) and added \benchmarksmetrics and \limitationsmitigations.

Then, at least two co-authors were assigned to each study type and guideline to refine and extend the content, followed by a peer review and refinement phase.
Afterwards, the first author reviewed all sections and discussed potential changes with the topic leads.
A requirement for co-authorship was that each author contributed or reviewed content and, most importantly, read the complete guidelines and confirmed that they stand behind all recommendations, not only their own contributions.

To select illustrative examples for each study type and guideline, co-author pairs independently searched for relevant papers using academic databases.
This initial pool was extended based on suggestions from other co-authors.
Identified papers were assessed by one author and subsequently reviewed by a second author; papers that did not add new insights beyond those already covered were not included.
Multiple review rounds were then carried out on the guidelines as a whole, during which selected papers were cross-checked by additional authors and, where necessary, further examples were added or replaced.
The examples are intended to be illustrative, not the result of a systematic review; the involvement of multiple authors at different stages helped mitigate individual selection bias.

After completing the internal review, we invited external experts in empirical SE to review the study types and guidelines.
During the review process of the \emph{Empirical Software Engineering} journal, the structure and content were improved based on reviewers' suggestions.
We self-assessed our guidelines against the quality dimensions of the \emph{SIGSOFT Empirical Standards}~\citep{ralph2021empiricalstandardssoftwareengineering}. This assessment is reported in the conclusion.

\section{Study Types}
\label{sec:study-types}

The SE community needs guidelines to establish common standards for designing and reporting empirical studies involving LLMs.
However, such guidelines must be tailored to different study types that each pose unique challenges.
Therefore, we created a taxonomy of study types to contextualize our recommendations.
We use the term \emph{study type} to refer to categories of LLM involvement in empirical research, rather than to research methodologies (e.g., experiments, case studies). A single empirical study may involve multiple such study types.
Each study type section starts with a \textbf{description}, followed by \textbf{examples} from the SE research community and beyond. The \textbf{advantages} and \textbf{challenges} of using LLMs are discussed in a summary subsection at the end of this section, synthesizing cross-cutting themes across all study types.
\ifpaper
See Table~\ref{tab:study-types} for an overview of study types.
\fi
\space%

\begin{table}[tb]
\centering
\caption{Overview of study types.}
\label{tab:study-types}
\begin{tabular}{@{}lll@{}}
\toprule
\textbf{Section} & \textbf{Title} & \textbf{Short Name} \\
\midrule
4.1   & \textbf{\hyperref[sec:llms-as-tools-for-software-engineering-researchers]{LLMs as Tools for Software Engineering Researchers}}                 &            \\
4.1.1 & \quad \hyperref[sec:llms-as-annotators]{LLMs as Annotators}                                                                                  & Annotators \\
4.1.2 & \quad \hyperref[sec:llms-as-judges]{LLMs as Judges}                                                                                          & Judges     \\
4.1.3 & \quad \hyperref[sec:llms-for-synthesis]{LLMs for Synthesis}                                                                                  & Synthesis  \\
4.1.4 & \quad \hyperref[sec:llms-as-subjects]{LLMs as Subjects}                                                                                      & Subjects   \\
\midrule
4.2   & \textbf{\hyperref[sec:llms-as-tools-for-software-engineers]{LLMs as Tools for Software Engineers}}                                           &            \\
4.2.1 & \quad \hyperref[sec:studying-llm-usage-in-software-engineering]{Studying LLM Usage in Software Engineering}                                  & Usage      \\
4.2.2 & \quad \hyperref[sec:llms-for-new-software-engineering-tools]{LLMs for New Software Engineering Tools}                                        & Tools      \\
4.2.3 & \quad \hyperref[sec:benchmarking-llms-for-software-engineering-tasks]{Benchmarking LLMs for Software Engineering Tasks}                      & Benchmarking \\
\bottomrule
\end{tabular}
\end{table}

\subsection{LLMs as Tools for Software Engineering Researchers}
\label{sec:llms-as-tools-for-software-engineering-researchers}

LLMs can serve as tools to help researchers conduct empirical studies.
They can automate various tasks such as data collection, pre-processing, and analysis.
For example, LLMs can perform qualitative coding on natural language text such as interview transcripts (\annotators), assess the quality of software artifacts (\judges), generate summaries of research papers (\synthesis), or simulate human behavior in empirical studies (\subjects). If LLMs could complete these tasks well enough, they would reduce the time and effort required to conduct a study. However, these applications each pose challenges to validity and reproducibility.
\space%

\studytypesubsection{LLMs as Annotators}
\label{sec:llms-as-annotators}

In the annotator role, LLMs perform qualitative coding---the annotation of natural language text such as requirements, interview transcripts, or open-ended survey responses---that researchers would otherwise do by hand.

\studytypeparagraph{Description}
Coding is a time-consuming manual process~\cite{DBLP:journals/ase/BanoHZT24}. LLMs can augment this process, suggest new codes, and label artifacts based on a pre-defined coding guide much faster than humans can~\cite{DBLP:conf/chi/HeHDRH24}. This covers both \emph{closed coding}, where the LLM labels artifacts against a predefined coding guide, and \emph{open coding}, where the LLM proposes new codes from the data. In closed coding, LLM labels can be assessed against those of human coders applying the same guide; in open coding, researchers review the resulting codebook for adequacy (e.g., level of abstraction, redundancies between codes).
The extent to which, or under what conditions, LLMs can perform these tasks \textit{effectively} remains an open research question~\cite{DBLP:conf/msr/AhmedDTP25}, and whether they should be used at all for reflexive qualitative analysis is itself contested~\cite{jowsey2025reject}. Indeed, measuring their effectiveness is practically and philosophically challenging. From an interpretivist philosophical perspective, one cannot measure the quality of analysis by comparing one (human or machine) analyst's work to another. From a realist perspective, triangulating across multiple human judges, LLM judges, and other data sources (e.g., whether a pull request is marked as having resolved an issue) improves confidence in the findings but does not prove that any one judge is valid.

\studytypeparagraph{Examples}
\citeauthor{Huang2023Enhancing} used multiple LLMs for joint annotation of mobile application reviews~\cite{Huang2023Enhancing}.
They used three models of comparable size with an absolute majority voting rule (i.e., a label is only accepted if it receives more than half of the total votes from the models). This approach slightly outperformed the best individual model tested. 
Meanwhile, \citeauthor{DBLP:conf/msr/AhmedDTP25} examined LLMs as annotators in SE research across five datasets, six LLMs, and ten annotation tasks~\cite{DBLP:conf/msr/AhmedDTP25}.
They found that inter-model agreement strongly correlates with human-model agreement; models performed poorly in tasks where humans also frequently disagreed. They proposed to use model confidence scores to identify specific samples that could be safely delegated to LLMs, potentially reducing human annotation effort without compromising inter-rater agreement.
\space%

\studytypesubsection{LLMs as Judges}
\label{sec:llms-as-judges}

In the judge role, LLMs rate or rank artifacts on quality criteria, in contrast to the qualitative coding tasks of \annotators.

\studytypeparagraph{Description}

As judges, LLMs rate software artifacts along quality criteria (e.g., code readability, adherence to coding standards, comment quality) or rank candidate solutions against such criteria. The scoring rubric is usually embedded in the prompt, with either a numerical scale or a binary verdict.

\studytypeparagraph{Examples}

\citeauthor{DBLP:conf/re/LubosFTGMEL24} used \href{https://www.llama.com/llama2/}{Llama-2} to evaluate the quality of software requirements statements~\cite{DBLP:conf/re/LubosFTGMEL24}.
They prompted the LLM with the text below, where the words in braces reflect the study parameters:

\begin{plain}
Your task is to evaluate the quality of a software requirement.
Evaluate whether the following requirement is {quality_characteristic}.
{quality_characteristic} means: {quality_characteristic_explanation}
The evaluation result must be: 'yes' or 'no'.
Request: Based on the following description of the project: {project_description}
Evaluate the quality of the following requirement: {requirement}.
Explain your decision and suggest an improved version.
\end{plain}

They evaluated LLM output against expert human judges and found moderate agreement for simple requirements and poor agreement for more complex requirements. In contrast, \citeauthor{DBLP:conf/kbse/WangALHTASL25} used an LLM to generate acceptance criteria for user stories, and provided a rubric to an LLM to judge the generated acceptance criteria on interpretable scales (0 to 4)~\cite{DBLP:conf/kbse/WangALHTASL25}.
\space%

\studytypesubsection{LLMs for Synthesis}
\label{sec:llms-for-synthesis}

In the synthesis role, LLMs integrate and interpret information from multiple sources to produce higher-level findings such as themes, patterns, or conceptual frameworks. They can also generate synthetic content (e.g., source code, bug-fix pairs, requirements) for downstream training or evaluation.

\studytypeparagraph{Description}

Unlike annotation (see \annotators), which focuses on categorizing or labeling individual data points, synthesis refers to the process of integrating and interpreting information from multiple sources to generate higher-level insights, identify patterns across datasets, and develop conceptual frameworks or theories. LLMs may be able to support synthesis tasks in SE research by processing and distilling information from qualitative data sources. 
Although synthesis in the preceding notion refers to abstraction and interpretation across multiple data sources, the term is sometimes also used to refer to generating synthetic content (e.g., source code, bug-fix pairs, or requirements) that is then used in downstream tasks to train, fine-tune, or evaluate existing models or tools.
In this case, the synthesis is done primarily using the LLM and its training data; the input is limited to basic instructions and examples.

\studytypeparagraph{Examples} 

Published examples of applying LLMs for synthesis in SE remain scarce. However, some recent work in other domains is instructive~\cite{DBLP:journals/ase/BanoHZT24}.
\citeauthor{DBLP:conf/wsese/BarrosANKKNB25} conducted a systematic mapping study on using LLMs for qualitative research~\cite{DBLP:conf/wsese/BarrosANKKNB25}. The included studies span fields such as healthcare, education, and cultural studies, where LLMs supported qualitative methods, such as grounded theory and thematic analysis, by aiding in pattern identification.
In SE, \citeauthor{DBLP:conf/wsese/LecaVSS25} explored how LLMs have been applied for qualitative data analysis (QDA) and proposed general strategies and guidelines for their application~\cite{DBLP:conf/wsese/LecaVSS25}. \citeauthor{DBLP:journals/corr/abs-2511-14528} complemented this perspective by studying the opportunities and limitations of introducing LLM-based support into QDA, and by formulating recommendations for embedding human–AI collaboration across the thematic analysis phases~\cite{DBLP:journals/corr/abs-2511-14528}. Building on these, subsequent work has proposed hybrid frameworks combining LLM support with human-led QDA. \citeauthor{DBLP:journals/corr/abs-2402-01386} designed an LLM-driven multi-agent system that integrates AI with human decision-making to automate qualitative data analysis methods~\cite{DBLP:journals/corr/abs-2402-01386}. Their system generated initial codes, developed themes, and summarized text. Similarly, \citeauthor{DBLP:journals/corr/abs-2510-18456} compared the performance of humans and LLMs in coding, theme development, definition, and refinement, creating guidelines for a hybrid-LLM framework~\cite{DBLP:journals/corr/abs-2510-18456}.
Finally, \citeauthor{DBLP:journals/corr/abs-2506-21138}'s work is an example of using LLMs to create synthetic datasets.
They presented an approach to generate synthetic requirements, showing that they \enq{can match or surpass human-authored requirements for specific classification tasks}~\cite{DBLP:journals/corr/abs-2506-21138}.
\space%

\studytypesubsection{LLMs as Subjects}
\label{sec:llms-as-subjects}

In the subject role, LLMs serve as virtual subjects, generating responses or behaviors that an empirical study would otherwise collect from human participants.

\studytypeparagraph{Description}

In empirical studies, data is collected from participants through methods such as surveys, interviews, or controlled experiments.
LLMs can serve as virtual \emph{subjects} by simulating human behavior and interactions. If LLMs can generate responses that approximate those of human participants, they could be valuable for research involving user interactions, collaborative coding environments, and software usability assessments~\cite{ZHAO2025101167}.
To achieve this, prompt engineering techniques are widely employed; for instance, the \textit{Personas Pattern}~\cite{DBLP:conf/naacl/KongZCLQSZWD24} involves tailoring LLM responses to align with predefined profiles or roles that emulate specific user archetypes.
To serve as virtual subjects, generated responses should be indistinguishable from human-produced texts, consistent with the attitudes and sociodemographic information of the conditioning context (e.g., junior vs.\ senior developers), naturally aligned with the form, tone, and content of the simulated scenario, and reflect patterns in relationships between ideas, demographics, and behavior observed in comparable human data~\cite{DBLP:journals/corr/abs-2209-06899}.

\studytypeparagraph{Examples}

\citeauthor{DBLP:journals/ipm/XuSRGPLSH24} compiled a list of ways LLMs can support social science research, some of which transfer to empirical SE research~\cite{DBLP:journals/ipm/XuSRGPLSH24}. For example, LLMs can emulate human responses and behaviors in simulated interviews and focus groups~\cite{DBLP:journals/ase/GerosaTSS24}. Similarly, \citeauthor{DBLP:conf/icse/BanoGH25} investigated biases in LLM-generated candidate profiles in SE recruitment processes~\cite{DBLP:conf/icse/BanoGH25}. They found biases favoring male candidates, lighter skin tones, and slim physiques, particularly for senior roles. LLMs may be able to simulate end-user feedback and behavior in usability studies, identify usability issues, and offer suggestions for improvement based on predefined user personas.
\space%

\subsection{LLMs as Tools for Software Engineers}
\label{sec:llms-as-tools-for-software-engineers}

LLM-based assistants have become a widely adopted tool for software engineers~\cite{stackoverflow2025survey}, supporting them in various tasks such as code generation and debugging.
Researchers have studied how software engineers use LLMs (\llmusage), developed new tools that integrate LLMs (\newtools), and benchmarked LLMs for software engineering tasks (\benchmarkingtasks).
\space%

\studytypesubsection{Studying LLM Usage in Software Engineering}
\label{sec:studying-llm-usage-in-software-engineering}

Studies of LLM usage examine how software engineers adopt and integrate LLM-based tools into their workflows.

\studytypeparagraph{Description}

LLM usage studies focus on real-world settings, outside the controlled conditions of benchmarks and experiments.
Researchers can observe software engineers' usage of LLM-based tools in the field, or study if and how they adopt such tools, their usage patterns, as well as perceived benefits and challenges.
Surveys, interviews, observational studies, or analysis of usage logs can provide insights into how LLMs are integrated into development processes, how they influence decision making, and what factors affect their acceptance and effectiveness. 
Such studies can inform improvements for existing LLM-based tools, motivate the design of novel tools, or derive best practices for LLM-assisted software engineering.
They can also uncover risks or deficiencies of existing tools.

\studytypeparagraph{Examples}

Based on a convergent mixed-methods study, \citeauthor{DBLP:journals/tosem/Russo24} found that early adoption of generative AI by software engineers is primarily driven by compatibility with existing workflows~\cite{DBLP:journals/tosem/Russo24}.
\citeauthor{DBLP:journals/pacmse/KhojahM0N24} investigated the use of ChatGPT (GPT-3.5) by professional software engineers in a week-long observational study~\cite{DBLP:journals/pacmse/KhojahM0N24}.
They found that most developers do not use the code generated by ChatGPT directly but instead use the output as a guide to implement their own solutions.
\citeauthor{DBLP:conf/csee/AzanzaPIG24}'s case study found that LLMs could \enq{enhance personalized, instant onboarding support; however, relying on proprietary external LLMs poses significant data privacy risks}~\cite{DBLP:conf/csee/AzanzaPIG24}. 
 \citeauthor{DBLP:conf/icsa/JahicS24} found that most participants from 15 software companies they surveyed had already adopted AI (especially ChatGPT) for SE tasks, but cited low-quality outputs, copyright issues, and the risk of proprietary code leaks as barriers to adoption~\cite{DBLP:conf/icsa/JahicS24}. 
Retrospective studies that analyze data generated while developers use LLMs can provide additional insights into human-LLM interactions.
For example, researchers can employ data mining methods to build large-scale conversation datasets, such as the DevGPT dataset introduced by \citeauthor{DBLP:conf/msr/XiaoTHM24}~\cite{DBLP:conf/msr/XiaoTHM24}.
Conversations can then be analyzed using quantitative~\cite{DBLP:conf/msr/RabbiCZI24} and qualitative~\cite{DBLP:conf/msr/MohamedPP24} analysis methods.
\space%

\studytypesubsection{LLMs for New Software Engineering Tools}
\label{sec:llms-for-new-software-engineering-tools}

In this role, LLMs serve as components of new tools that assist software engineers (e.g., with code comprehension or test generation) or as autonomous agents that perform multi-step tasks on their behalf.

\studytypeparagraph{Description}

New LLM-based tools support software engineers in their daily tasks, such as code comprehension~\cite{DBLP:conf/chi/YanHWH24} and test case generation~\cite{DBLP:journals/tse/SchaferNET24}.
One way of integrating LLM-based tools into software engineers' workflows is using GenAI agents.
Unlike traditional LLM-based tools, these agents are capable of acting autonomously and proactively, are often tailored to meet specific user needs (e.g., via context files or domain-specific tools), and can interact with external environments (e.g., file systems, shells, or web APIs)~\cite{wiesinger2025agents,yang2025code}.
From an architectural perspective, GenAI agents can be implemented in various ways~\cite{wiesinger2025agents}, but at their core they run a control loop around the LLM (observe $\rightarrow$ inspect $\rightarrow$ choose $\rightarrow$ act)~\cite{raschka2026components}.
Each chosen action (e.g., a tool call) produces a result that the agent feeds back into the next iteration, until the task is complete.
\emph{CoALA}~\cite{DBLP:journals/tmlr/SumersYN024} offers a conceptual framework for organizing such agents.
For coding agents specifically, \citeauthor{raschka2026components} identifies building blocks such as the repository context gathered before each call, the constructed prompt and its tool definitions, structured session memory, and delegation to bounded subagents~\cite{raschka2026components}.
Because these architectures vary, researchers can test and compare them to study how design choices affect downstream performance.

\studytypeparagraph{Examples}

\citeauthor{DBLP:conf/chi/YanHWH24} proposed \emph{IVIE}, a tool integrated into the VS Code graphical interface that generates and explains code using LLMs~\cite{DBLP:conf/chi/YanHWH24}.
The authors focused more on the presentation, providing a user-friendly interface to interact with the LLM. 
\citeauthor{DBLP:journals/tse/SchaferNET24} presented a large-scale empirical evaluation on the effectiveness of LLMs for automated unit test generation~\cite{DBLP:journals/tse/SchaferNET24}.
They presented \emph{TestPilot}, a tool that implements an approach in which the LLM is provided with prompts that include the signature and implementation of a function under test, along with usage examples extracted from the documentation.
\citeauthor{DBLP:conf/icsm/RichardsW24} introduced a preliminary GenAI agent designed to assist developers in understanding source code by incorporating a reasoning component grounded in the theory of mind~\cite{DBLP:conf/icsm/RichardsW24}.
\citeauthor{DBLP:conf/icse-seip/TakerngsaksiriPTTZJLCCW25} presented \emph{HULA}, a multi-agent system deployed in Atlassian JIRA that lets engineers refine LLM-generated coding plans and source code, and reported acceptance and modification rates from real users~\cite{DBLP:conf/icse-seip/TakerngsaksiriPTTZJLCCW25}.
\space%

\studytypesubsection{Benchmarking LLMs for Software Engineering Tasks}
\label{sec:benchmarking-llms-for-software-engineering-tasks}

Benchmarking studies measure LLM performance on standardized SE tasks against reference outputs and shared metrics.

\studytypeparagraph{Description}

In a benchmark, the reference outputs serve as ground truth, and the metrics measure how well an LLM's outputs match them. Typical tasks include code generation, code summarization, code completion, and code repair~\cite{yang2025code}, but also natural language processing tasks such as anaphora resolution (i.e., the task of identifying the referring expression of a word or phrase occurring earlier in the text). 
Metrics may include general metrics for text generation, such as \emph{ROUGE}, \emph{BLEU}, or \emph{METEOR}~\cite{DBLP:journals/tosem/HouZLYWLLLGW24}, or task-specific metrics, such as \emph{CodeBLEU} for code generation. Benchmarking requires high-quality reference datasets.

\studytypeparagraph{Examples}

In SE, benchmarking may include evaluating an LLM's ability to produce accurate and reliable outputs for a given input (usually a task description, possibly accompanied by data obtained from curated real-world projects or from synthetic SE-specific datasets). \emph{RepairBench}~\cite{DBLP:conf/llm4code/SilvaM25}, for example, contains 574 buggy Java methods and their corresponding fixed versions, which can be used to evaluate the performance of LLMs in code repair tasks. It uses the \emph{Plausible@1} metric (i.e., the probability that the first generated patch passes all test cases) and the \emph{AST Match@1} metric (i.e., the probability that the abstract syntax tree of the first generated patch matches the ground truth patch).
\emph{SWE-Bench}~\cite{DBLP:conf/iclr/JimenezYWYPPN24} is a more generic benchmark that contains 2,294 SE Python tasks extracted from GitHub pull requests.
To score the LLM's task performance, the benchmark validates whether the generated patch successfully compiles and calculates the percentage of passed test cases. Meanwhile, \emph{HumanEval}~\cite{DBLP:journals/corr/abs-2107-03374} is often used to assess code generation. 
\space%

\subsection{Advantages and Challenges}
\label{sec:advantages-challenges}

Using LLMs in empirical SE research, whether as tools for researchers or for software engineers, offers several potential advantages but also raises fundamental challenges that cut across the study types described above.

\studytypeparagraph{Advantages}

The primary advantage of using LLMs for research tasks is \emph{speed, cost reduction, and scalability}. LLMs can annotate, judge, synthesize, and simulate faster and at lower cost than human researchers, with studies showing cost reductions of 50--96\% on various natural language tasks~\cite{DBLP:conf/emnlp/WangLXZZ21} (e.g., \citeauthor{DBLP:conf/chi/HeHDRH24} found that GPT-4 annotation required only two days and 122.08 USD compared to several weeks and 4,508 USD for a comparable MTurk pipeline~\cite{DBLP:conf/chi/HeHDRH24}). Similarly, augmenting or replacing human participants with LLM-generated virtual participants would reduce recruitment effort~\cite{DBLP:conf/vl/Madampe0HO24}. This efficiency unlocks scalability: qualitative research traditionally does not scale well to large samples, but LLMs let researchers process more text than human-only coding allows and support larger judgment datasets.

LLMs can also \emph{automate} tasks such as coding qualitative data, assessing artifact quality, and generating summaries, reducing cognitive demands and resources required for qualitative research. Some studies suggest that LLMs may improve \emph{consistency}: ChatGPT's accuracy exceeded crowd workers by approximately 25\%, and LLMs can achieve higher inter-rater agreement than crowd workers and trained annotators~\cite{DBLP:journals/corr/abs-2303-15056}. However, both human and LLM judges exhibit systematic biases, such as preferring outputs with fake citations or rich formatting~\cite{DBLP:conf/emnlp/ChenCLJW24}, and no compelling evidence supports claims that LLMs are less biased than human annotators.

LLMs could potentially provide \emph{access to otherwise-inaccessible research contexts}. \textit{If} virtual participants' behavior is sufficiently similar to human participants, LLMs could access underrepresented and hard-to-reach populations, strengthen generalizability and inclusiveness, impute missing data (see \synthesis), and enable research that is ethically problematic with real humans (e.g., questions that would force a human to relive past trauma do not harm an LLM).

Studying real-world LLM-based tools allows researchers to \emph{understand the state of practice}, uncovering usage patterns, adoption rates, and contextual factors, and \emph{generate hypotheses} about how LLMs affect developer productivity, collaboration, and decision-making. From an engineering perspective, developing LLM-based tools \emph{requires less task-specific engineering} than traditional SE approaches such as static analysis or symbolic execution, because a single model can handle diverse inputs without language-specific parsers or hand-crafted rules. Good benchmarks provide \emph{standardized evaluation and model comparison}, reducing research effort and supporting open science by providing common ground for sharing data and results. Benchmarks built for specific SE tasks can help identify LLM weaknesses and support optimization and fine-tuning.

\studytypeparagraph{Challenges}

The stochastic nature of LLM responses, where identical prompts may yield different outputs, \emph{undermines replicability} across all study types, complicating interpretation of experimental results and test-retest reliability. More broadly, \emph{reliability} is a persistent concern: Like any measurement instrument, \judges and \annotators should exhibit validity, test-retest reliability, inter-rater reliability, minimal error, and measurement invariance~\cite{DBLP:books/sp/24/RalphKAA24}. However, LLMs show substantial variability depending on the dataset and task~\cite{DBLP:journals/corr/abs-2306-00176, DBLP:journals/corr/abs-2406-18403}. They are sensitive to prompt variations~\cite{DBLP:journals/corr/abs-2304-11085} and option order~\cite{DBLP:conf/naacl/PezeshkpourH24}, can behave differently when reviewing their own outputs~\cite{DBLP:conf/nips/PanicksseryBF24}, and can be unreliable for high-stakes labeling~\cite{DBLP:conf/chi/Wang0RMM24}. Crucially, \emph{reliability does not imply validity}: a reliable LLM might be reliably inaccurate~\cite{DBLP:conf/coling/ZhouC024}. For tasks with no single correct answer, the statistical framework pushes outputs toward the most likely answer, which may not be the best one~\cite{DBLP:journals/corr/abs-2510-01171, DBLP:journals/corr/abs-2310-06452}. See \modelversion and \limitationsmitigations for reporting guidance on replicability and reliability.

LLMs and LLM-based tools \emph{evolve rapidly}, and so do practitioners' workflows around them. This combination complicates longitudinal comparisons and can quickly render study findings obsolete.
Findings tied to a specific model version may not transfer to later releases even within the same model family, complicating reproducibility and longitudinal comparisons.

The prevalence of \emph{proprietary tools and opaque training data} limits researchers' ability to assess and mitigate biases, and enables benchmark contamination~\cite{DBLP:journals/corr/abs-2410-16186}. LLMs exhibit \emph{bias} in multiple forms: tendencies to overestimate certain labels~\cite{DBLP:journals/corr/abs-2304-10145}, well-documented fairness issues~\cite{DBLP:journals/coling/GallegosRBTKDYZA24}, and the potential to reinforce prejudices. When simulating human participants, LLMs are \enq{likely to both misportray and flatten the representations of demographic groups}~\cite{DBLP:journals/natmi/WangMD25}. See \openllm and \limitationsmitigations for mitigation strategies.

\emph{Evaluation difficulty} is inherent in studying LLM-based tools and benchmarks. Benchmarks may lack construct validity~\cite{DBLP:books/sp/24/RalphKAA24}, usually do not capture the full complexity of software engineering work~\cite{Chandra2025benchmarks}, and may lead to \emph{overconfidence} and overfitting~\cite{DBLP:journals/corr/abs-2412-03597}. LLMs are also susceptible to adversarial manipulation where semantics-preserving changes may deceive them into accepting flawed artifacts~\cite{DBLP:journals/corr/abs-2510-24367}. See \benchmarksmetrics for detailed guidance on benchmark design, including~\citeauthor{cao2025should}'s guidelines for coding task benchmarks~\cite{cao2025should}.

A fundamental challenge is \emph{philosophical and methodological incongruence} with qualitative research. In a recent open letter, 419 experienced qualitative researchers argued \enq{that analytic approaches such as reflexive thematic analysis are human research practices requiring a subjective, positioned, and reflexive researcher and therefore the use of GenAI in such approaches is not methodologically congruent}~\cite{jowsey2025reject}. LLMs lack the capacity for genuinely reflexive qualitative analysis because they operate on statistical prediction without understanding the meaning of the data being analyzed~\cite{DBLP:journals/corr/abs-2306-13298, DBLP:journals/corr/abs-2510-18456}. Effective use requires structured prompts and careful human oversight~\cite{DBLP:conf/wsese/BarrosANKKNB25}. See \humanvalidation and \limitationsmitigations for guidance.

There is \emph{insufficient evidence} for the effectiveness of LLMs in most research roles. No compelling evidence exists that LLMs can accurately simulate human participants~\cite{DBLP:journals/corr/abs-2508-06950} or reliably judge most relevant properties of SE artifacts. Gathering such evidence for each specific usage may be quite difficult~\cite{DBLP:journals/ais/HardingDLL24}. LLMs may be better suited for augmenting rather than replacing human researchers~\cite{DBLP:conf/emnlp/WangLXZZ21, DBLP:conf/chi/HeHDRH24}, but even then, limited evidence exists that augmentation increases \emph{effectiveness} as well as \emph{efficiency}. When LLM outputs are incorrect, they can negatively \emph{affect human judgment}~\cite{DBLP:conf/www/HuangKA23a}. See \humanvalidation for mitigation strategies.

Majority voting across multiple outputs improves reliability~\cite{DBLP:journals/corr/abs-2304-11085} but increases \emph{cost and environmental impact}. While open models are available, the most capable ones require substantial hardware; relying on \emph{cloud-based APIs} introduces concerns related to data privacy, security, and replicability. See \limitationsmitigations for sustainability considerations.

Field study findings face \emph{generalizability} challenges because outcomes may be highly sensitive to individual differences, usage patterns, goals, and contexts. Field studies must be \enq{dependable}~\cite{Sullivan2011-ub} beyond traditional validity criteria, which complicates methodology; see \limitationsmitigations for a detailed threat taxonomy.
\space%

\section{Guidelines}
\label{sec:guidelines}

A primary goal of our guidelines is to \emph{enable reproducibility and replicability} of empirical SE studies involving LLMs. As repeating LLM-focused research to verify results lies somewhere between the ACM's definitions of reproducibility (different team, same research artifacts) and replicability (different team, different research artifacts)~\cite{acm2020artifactreview} due to potential model changes and incomplete research artifacts (e.g., unreported prompts, configurations, or seeds), we follow \citeauthor{DBLP:journals/corr/abs-2510-25506} and use the terms interchangeably~\cite{DBLP:journals/corr/abs-2510-25506}. While previous guidelines regarding open science and empirical studies still apply, LLM-specific characteristics (e.g. inherent non-determinism~\cite{DBLP:conf/naacl/SongWLL25, YuanNondeterminismLLMInference}, opaque and proprietary models) present additional replicability challenges, which, in turn, demand new guidance.

Each guideline below begins with a brief \textbf{summary}, followed by its \textbf{rationale}, \textbf{recommendations}, \textbf{examples}, \textbf{benefits}, and \textbf{challenges}, with links to the relevant \textbf{study types}.
The \emph{rationale} articulates the underlying principle, i.e., why the guideline matters, while the \emph{recommendations} provide concrete, actionable practices.
\ifpaper
Table~\ref{tab:rationale-recommendations} in the appendix provides a compact overview of this mapping.
\fi

The guidelines further contain \textbf{advice for reviewers} and close with a \textbf{see also} list linking related guidelines.
Broadly, reviewers should use our guidelines to help them interrogate the extent to which a manuscript's authors have done what was practically possible to improve reproducibility. We must neither accept research absent reasonable efforts to improve reproducibility, nor reject research for failing to obtain an impossible goal. We borrow this principle of \enq{reasonable efforts} from the SIGSOFT Empirical Standards~\cite{ralph2021empiricalstandardssoftwareengineering}, where it applies to methodological rigor more generally. Of course, papers should acknowledge their limitations, but to determine whether these limitations are reasonable, reviewers should ask ``have the authors done what they could to minimize limitations?'' Our guidelines attempt to capture what \enq{reasonable efforts} practically means for LLM-based studies.

To distinguish essential criteria from recommendations, our guidelines use two tiers.
A~\must criterion is a \emph{requirement}. Studies that intend to follow these
guidelines are expected to meet all \must criteria.
A~\should criterion represents a desired practice that strengthens a study's rigor or transparency.
However, there may be valid reasons to deviate in particular circumstances (e.g., resource constraints, inapplicability to a specific study context or type).
Nonetheless, authors should briefly justify any deviation from a \should criterion that could compromise a study's validity or reproducibility.
The distinction between \must and \should also reflects \emph{what} the criterion mandates, not only its severity.
A \must criterion is typically a disclosure obligation such as reporting model versions, publishing prompts, describing architectures, and discussing limitations, so that readers can independently evaluate the choices authors made.
A \should criterion is typically a methodological recommendation such as which baselines to include and which validation strategies or statistical analyses to apply.
\ifpaper
Table~\ref{tab:guidelines} lists our eight guidelines.
Table~\ref{tab:guideline-matrix} in the appendix provides an overview of how each guideline applies to each study type.
Cells marked \iconM indicate that the guideline is a \must requirement for the study type and cells marked \iconS indicate that the guideline \should be followed.
Cells marked -- indicate that the guideline is typically not applicable or not feasible for the study type.
\fi

The following sections indicate which information we expect researchers to report, and whether it should be in the \paper or \supplementarymaterial. Where a publication venue's page limits hinder reporting all expected elements in the \paper, it is better to report essential information in the \supplementarymaterial than not at all.
The \supplementarymaterial \should be published according to the \emph{ACM SIGSOFT Open Science Policies}~\citep{daniel_graziotin_2024_10796477}.
\space%

\begin{table}[tb]
\centering
\setlength{\tabcolsep}{5pt}
\caption{Overview of guidelines.}
\label{tab:guidelines}
\begin{tabular}{@{}lll@{}}
\toprule
\textbf{Section} & \textbf{Title} & \textbf{Short Name} \\
\midrule
\ref*{sec:declare-llm-usage-and-role}                            & \hyperref[sec:declare-llm-usage-and-role]{Declare LLM Usage and Role}                                          & Declare Usage    \\
\ref*{sec:report-model-version-configuration-and-customizations} & \hyperref[sec:report-model-version-configuration-and-customizations]{Report Model Version, Configuration, and Customizations} & Model Version    \\
\ref*{sec:report-system-and-prompt-design}                       & \hyperref[sec:report-system-and-prompt-design]{Report System and Prompt Design}                                & Design           \\
\ref*{sec:report-session-traces}                                 & \hyperref[sec:report-session-traces]{Report Session Traces}                                                    & Traces           \\
\ref*{sec:use-suitable-baselines-benchmarks-and-metrics}         & \hyperref[sec:use-suitable-baselines-benchmarks-and-metrics]{Use Suitable Baselines, Benchmarks, and Metrics}  & \begin{tabular}[t]{@{}l@{}}Benchmarks\\\& Metrics\end{tabular} \\
\ref*{sec:use-an-open-llm-as-a-baseline}                         & \hyperref[sec:use-an-open-llm-as-a-baseline]{Use an Open LLM as a Baseline}                                    & Open LLM         \\
\ref*{sec:use-human-validation-for-llm-outputs}                  & \hyperref[sec:use-human-validation-for-llm-outputs]{Use Human Validation for LLM Outputs}                      & Human Validation \\
\ref*{sec:report-limitations-and-mitigations}                    & \hyperref[sec:report-limitations-and-mitigations]{Report Limitations and Mitigations}                          & Limitations      \\
\bottomrule
\end{tabular}
\end{table}

\subsection{Declare LLM Usage and Role}
\label{sec:declare-llm-usage-and-role}

\vspace{0.5\baselineskip}
\begin{framed}
\summary: %
Researchers \must disclose any use of LLMs to support empirical studies, specifying which LLM was used, how it was used, and where in the research process it was employed. This disclosure \should appear in a suitable section of the \paper. They \should report the exact purpose, the tasks that were automated, and the expected benefits in the \paper. When the LLM is central to the study, the declaration \should be prominent and detailed in the methodology section; for tangential uses, a brief statement in the methodology section suffices.
\end{framed}

\guidelinesubsubsection{Rationale}

Transparency about LLM involvement is a prerequisite for informed assessment of a study's scope, limitations, and potential biases.
Without explicit disclosure, readers cannot evaluate how the LLM's characteristics may have influenced the research process or its outcomes.

\guidelinesubsubsection{Recommendations}

When conducting any kind of empirical study involving LLMs, researchers \must clearly declare that an LLM was used (see \scope for what we consider relevant research support).
This \should be done in a suitable section of the \paper, for example, in the introduction or research methods section; \citeauthor{cheng2025writing} argue specifically for the methods section, since acknowledgments are easily missed at the end of a paper~\cite{cheng2025writing}.
The ACM Policy on Authorship requires authors to describe in the methods section any use of AI in the research itself~\cite{ACM2026}.

Beyond generic declarations, researchers \should report the exact purpose of using an LLM in a study, the tasks it was used to automate, and the expected benefits in the \paper.
A sufficient declaration specifies not only \emph{that} an LLM was used, but also \emph{which} LLM (name and version), \emph{how} it was used (e.g., as an annotator, code generator, or judge), and \emph{where} in the research process it was employed (e.g., data collection, analysis, or synthesis).

When the LLM is central to the study (e.g., as the main tool being evaluated or as a core component of the research method), the declaration \should be prominent and detailed, appearing in the methodology section with cross-references to the specific guidelines that apply (e.g., Sections \modelversion, \design, and \traces).
When the LLM's role is more tangential (e.g., used for a single preprocessing step), a brief but explicit statement in the methodology section is sufficient.
When a study assigns multiple distinct roles to LLMs (e.g., one model generates evaluation data while another scores outputs), each role \should be declared separately.
In each case, the disclosure \must be specific enough for readers to assess how the LLM's involvement may affect the study's validity and reproducibility.

\guidelinesubsubsection{Examples}

The \emph{ACM Policy on Authorship} requires reporting AI-generated artifacts such as code, datasets, and figures where they underlie a study's conclusions~\cite{ACM2026}.
Reporting in the methods section keeps this disclosure visible under double-blind review, where end-of-paper acknowledgments are often removed.
An example of an LLM disclosure beyond writing support can be found in a recent paper by \citeauthor{DBLP:conf/re/LubosFTGMEL24}~\cite{DBLP:conf/re/LubosFTGMEL24}, in which they write in the methodology section:

\begin{quote}
\it
``We conducted an LLM-based evaluation of requirements utilizing the Llama 2 language model with 70 billion parameters, fine-tuned to complete chat responses...''
\end{quote}

\noindent A more contemporary declaration could similarly state:

\begin{quote}
\it
``We used Claude Opus 4.7 via the Anthropic API to synthesize themes from interview transcripts, with all prompts and conversation logs published as supplementary material.''
\end{quote}

\citeauthor{DBLP:journals/corr/abs-2601-11895}'s DevBench paper illustrates separate disclosure of multiple LLM roles: \emph{GPT-4o} generated the synthetic benchmark instances, nine models (including \emph{Claude 4 Sonnet}, \emph{GPT-4.1}, \emph{DeepSeek-V3}, and \emph{Ministral-3B}) were the evaluation subjects, and \emph{o3-mini} was the LLM judge that scored completions for relevance and helpfulness~\cite{DBLP:journals/corr/abs-2601-11895}.

\guidelinesubsubsection{Benefits}

Transparency in the use of LLMs helps other researchers understand the context and scope of the study, supporting interpretation and comparison of the results.
Realizing these benefits requires reporting the LLM's exact role and version (see \modelversion).

\guidelinesubsubsection{Challenges}

Declaring LLM usage requires only a brief statement and no additional experiments, making compliance straightforward.
One challenge might be authors' reluctance to disclose LLM usage for valid use cases, because they fear that AI-generated content makes reviewers think that the authors' work is less original.
In fact, there is evidence suggesting that AI disclosure can negatively affect trust in authors~\cite{SCHILKE2025104405}.
However, the \emph{ACM Policy on Authorship} requires disclosure of AI used in the research itself, not AI used only to assist with writing~\cite{ACM2026}.
Our guidelines focus on such use (see \scope), not on proofreading or writing support.

\guidelinesubsubsection{Study Types}

Researchers \must follow this guideline for all study types.
The specific focus of the declaration varies by study type.
For \annotators, \judges, \synthesis, and \subjects, researchers \must declare the specific role assigned to the LLM (e.g., annotator, judge, synthesizer, or simulated participant).
For \llmusage, researchers \must clarify which LLM(s) the observed participants used and under which conditions.
For \newtools, researchers \must declare the LLM's role within the tool architecture and its contribution to the tool's functionality.
For \benchmarkingtasks, researchers \must declare which LLMs were benchmarked and for which tasks.

\guidelinesubsubsection{Advice for Reviewers}

The most common problem with disclosure is incompleteness or vagueness about how the LLM was used. If the paper says ``we used LLM X to help with task Y'' without specifying how, reviewers should request clarification. Such requests are typically minor revisions unless the missing details may reveal methodological problems.

If undisclosed LLM use is suspected, the reviewer should consult their editor or program chair. When the evidence is conclusive, the key question is the degree to which undisclosed use affects the study's contribution, ranging from negligible (e.g., word choice in a single sentence) to severe (e.g., generating the reported data or results).

\guidelinesubsubsection{See Also}
\begin{itemize}[label=$\rightarrow$]
    \item \seemodelversion: The disclosure is incomplete without naming the specific model and version.
    \item \seedesign: When the LLM lives inside a tool or agent, authors must also describe that tool's architecture and prompts.
    \item \seetraces: Session traces show what the LLM did during the study.
\end{itemize}

\subsection{Report Model Version, Configuration, and Customizations}
\label{sec:report-model-version-configuration-and-customizations}

\vspace{0.5\baselineskip}
\begin{framed}
\summary: %
Researchers \must report the exact LLM model or tool version, configuration, and date of study execution in the \paper. When using quantized models, researchers \should report the quantization level and method. For fine-tuned models, they \must describe the fine-tuning goal, dataset, and procedure in the \paper. Researchers \should include default parameters, explain model choices, compare base- and fine-tuned model using suitable metrics and benchmarks, and share fine-tuning data and weights as \supplementarymaterial (or alternatively justify in the \paper why they cannot share them).
\end{framed}

\guidelinesubsubsection{Rationale}

LLMs and LLM-based tools are frequently updated, and configuration parameters such as temperature or seed values affect content generation. 
This guideline focuses on documenting the \emph{model-specific} aspects of empirical studies involving LLMs, concentrating on the models themselves, their version, configuration parameters, and customizations (e.g., fine-tuning). While the \design section addresses system-level integration and the authored artifacts (including prompts) that the model uses on each call, the information outlined here is essential for reproducibility whenever an LLM is involved.

\guidelinesubsubsection{Recommendations}

Researchers \must document in the \paper which model or tool version they used in their study, along with the date of study execution and the parameters they configured that affect output generation.
Since default values might change over time, researchers \should report all configuration values, even if they used the defaults.
Checksums and fingerprints \should be reported since they identify specific versions and configurations.
Depending on the study context, other properties such as the context window size (number of tokens) \should be reported.
When using quantized models, researchers \should report the quantization level (e.g., 4-bit, 8-bit) and method (e.g., GPTQ or AWQ), as different quantization approaches produce different outputs, affecting both output quality and reproducibility.
Researchers \should motivate in the \paper why they selected certain models, versions, and configurations.
Reasons may be monetary, technical, or methodological (e.g., planned comparison to previous work).
Depending on the specific study context, additional information regarding the experiment or tool architecture \should be reported.

A common customization approach for existing LLMs is fine-tuning.
If a model was fine-tuned, researchers \must describe the fine-tuning goal (e.g., improving the performance for a specific task), the fine-tuning procedure (e.g., full fine-tuning versus Low-Rank Adaptation (LoRA), selected hyperparameters, loss function, learning rate, and batch size), and the fine-tuning dataset (e.g., data sources, the preprocessing pipeline, dataset size) in the \paper.
Researchers \should either share the fine-tuning dataset as part of the \supplementarymaterial or explain in the \paper why the data cannot be shared (e.g., because it contains confidential or personal data that could not be anonymized).
The same applies to the fine-tuned model weights.
Suitable benchmarks and metrics \should be used to compare the base model with the fine-tuned model.

In summary, researchers \must report in the \paper at minimum (1) the exact model or tool name and version, (2) all parameters they configured that affect output generation, (3) the date of study execution, and, for fine-tuned models, (4) the fine-tuning goal, dataset characterization, approach (e.g., full fine-tuning vs.\ LoRA), and hyperparameters. Beyond these requirements, researchers \should additionally report default parameter values, checksums or fingerprints, model properties relevant to the study (e.g., context-window size), and quantization level and method where applicable. For fine-tuned models, they \should also share the dataset and model weights as \supplementarymaterial (unless legal or privacy constraints prevent disclosure) and report validation metrics and benchmarks.

Commercial models (e.g., GPT-5) or LLM-based tools (e.g., ChatGPT) might not give researchers access to all required information.
For these tools, researchers \should report what is available and openly acknowledge limitations that hinder reproducibility.

\guidelinesubsubsection{Examples}

Based on the documentation that OpenAI and Azure provide~\cite{OpenAI25, Azure25}, researchers might, for example, report:

\begin{quote}
\it
 ``We integrated a  \texttt{gpt-4} model in version \texttt{0125-Preview} via the Azure OpenAI Service, and configured it with a temperature of 0.7, top\_p set to 0.8, a maximum token length of 512, and the  seed value \texttt{23487}.
 We ran our experiment on 10th January 2025. The system fingerprint was \texttt{fp\_6b68a8204b}.
\end{quote}

\citeauthor{DBLP:conf/icse/KangYY23} provide a similar statement in their paper on exploring LLM-based bug reproduction~\cite{DBLP:conf/icse/KangYY23}:

\begin{quote}
\it
``We access OpenAI Codex via its closed beta API, using the code-davinci-002 model. For Codex, we set the temperature to 0.7, and the maximum number of tokens to 256.''
\end{quote}

Our guidelines additionally recommend reporting a checksum/fingerprint and exact dates; otherwise, this example is close to our recommendations.

\citeauthor{DBLP:conf/icsa/DharVV24} assessed whether LLMs can generate architectural design decisions~\cite{DBLP:conf/icsa/DharVV24}, detailing the system architecture and the LLM's role within it.
They provide information on the fine-tuning approach and datasets, including the source of architectural decision records, preprocessing methods, and data selection criteria.

For self-hosted models, the \supplementarymaterial can become a true replication package. For example, for models provisioned using \href{https://ollama.com/library/}{ollama}, one can report the specific tag and checksum, e.g., \emph{``llama3.3, tag 70b-instruct-q8\_0, checksum d5b5e1b84868.''}
Given suitable hardware, running the model is then as easy as executing the following command:\\
\texttt{ollama run llama3.3:70b-instruct-q8\_0}

\guidelinesubsubsection{Benefits}

Reporting the model version, configuration, and date of study execution is a prerequisite for the verification and replication of LLM-based studies. While LLMs are inherently non-deterministic, this cannot excuse dismissing reproducibility. Although exact reproducibility is hard to achieve, the recommendations above help researchers come as close as possible to that standard.

\guidelinesubsubsection{Challenges}

Different model providers and modes of operating the models allow for varying degrees of information.
For example, OpenAI provides a model version and a system fingerprint describing the backend configuration, which can also influence the output.
However, the fingerprint is intended only to detect changes in the model or its configuration; one cannot go back to a certain fingerprint.
As a beta feature, OpenAI lets users set a seed parameter to receive \enq{(mostly) consistent output}~\cite{OpenAI23}.
However, the seed value does not allow for full reproducibility and the fingerprint changes frequently. 
Although, as motivated above, open models substantially simplify re-running experiments, they also come with challenges in terms of reproducibility, as generated outputs can be inconsistent despite setting the temperature to 0 and using a seed value (see \href{https://github.com/ollama/ollama/issues/5321}{GitHub issue for Llama3}).
Setting the temperature to 0 configures greedy decoding (always selecting the most probable next token), which minimizes output variability but can degrade quality by producing repetitive text and missing higher-quality responses~\cite{DBLP:conf/iclr/HoltzmanBDFC20}.

Even with a temperature of 0, full determinism is rarely guaranteed. Floating-point arithmetic on GPUs causes slight numerical differences that cascade into divergent token selections~\cite{YuanNondeterminismLLMInference}, and Sparse Mixture-of-Experts routing amplifies this effect~\cite{Chann2023}. Silent backend changes in commercial APIs produce different outputs over time~\cite{DBLP:journals/corr/abs-2307-09009}, and even self-hosted open models with identical settings do not always yield consistent outputs~\cite{DBLP:journals/corr/abs-2408-04667}.
Researchers \shouldnot treat a temperature of 0 as a guarantee of reproducibility, but as one measure among several, including fixed seed values~\cite{OpenAI23}, system fingerprints, and archiving of raw outputs.
When a temperature of 0 is chosen primarily for reproducibility, this motivation \should be stated explicitly, along with an acknowledgment of its potential impact on output quality.

\guidelinesubsubsection{Study Types}

This guideline \must be followed for all study types for which the researcher has access to (parts of) the model's configuration.
They \must always report the configuration that is visible to them, acknowledging the reproducibility challenges of commercial tools and models that are offered as-a-service.
Depending on the specific study type, researchers \should provide additional information on the system and prompt design (see \design), session traces (see \traces), and specific limitations and mitigations (see \limitationsmitigations).

For example, when \llmusage by focusing on commercial tools such as ChatGPT or GitHub Copilot, researchers \must be as specific as possible in describing their study setup.
The configured model name, version, and the date of study execution \must always be reported.
See \design for prompt reporting and \traces for interaction logs.

For \annotators, \judges, and \synthesis, researchers \must report the model configuration used for the respective annotation, judging, or synthesis tasks, including temperature and other sampling parameters that affect output variability.
For \subjects, researchers \must report any persona-related configuration settings and parameters that configure the simulated behavior.
For \newtools, researchers \must report the configuration for each model integrated in the tool's architecture, including any model-specific parameter choices.
For \benchmarkingtasks, researchers \must report the configuration for all benchmarked models to enable fair cross-model comparisons.

\guidelinesubsubsection{Advice for Reviewers}

Missing version, configuration, or parameter information is typically a minor revision request. Before concluding that information is absent, reviewers should check appendices and supplementary materials, as details are sometimes reported there rather than in the main text. Rejection over missing details is rarely warranted unless the omissions obscure deeper methodological problems.

\guidelinesubsubsection{See Also}
\begin{itemize}[label=$\rightarrow$]
    \item \seedesign: Beyond the model itself, authors must also document the architecture and prompts that use it.
    \item \seetraces: Session traces show what the reported model and configuration produced at runtime.
    \item \seebenchmarksmetrics: Benchmark comparisons require identifying the exact model version under test.
    \item \seeopenllm: An open model gives full version visibility, which some commercial products do not.
    \item \seelimitationsmitigations: Hidden or shifting commercial versions become reproducibility threats in their own right.
\end{itemize}

\subsection{Report System and Prompt Design}
\label{sec:report-system-and-prompt-design}

\vspace{0.5\baselineskip}
\begin{framed}
\summary: %
Researchers \must describe in the \paper the full architecture of LLM-based tools, from standalone uses to agentic systems, including the LLM's role and its interactions with other components. Hosting and access \should be described; for time-sensitive measurements, researchers \must clarify whether local infrastructure or cloud services were used. Researchers \must publish all prompts as \supplementarymaterial, including prompt templates with representative instances and the dynamic generation process where applicable, and \must include representative examples in the \paper; sensitive content \must be anonymized. If full prompt disclosure is not feasible, summaries or examples \should be provided. The prompting strategy and prompt reuse across models and configurations \must be specified, and how the prompts were developed \should be described; for few-shot prompts, how the examples were selected \must be explained in the \paper. Researchers \must describe the configuration mechanisms used (e.g., context files, skills, subagents) and summarize in the \paper which tools were exposed; all configuration artifacts and the complete tool catalog (schemas, definitions, Model Context Protocol (MCP) servers) \should be published as \supplementarymaterial. For agentic systems, researchers \must specify the agents' roles, reasoning frameworks, and communication flows; where external tools are used, the model's reasoning, tool calls, and interactions with users or the environment \must be reported separately. For retrieval-augmented generation (RAG) and similar methods, researchers \must describe how external data was retrieved, stored, and integrated. Where legally possible, the implementation \should be open-sourced; non-disclosed proprietary components \must be acknowledged as reproducibility limitations.
\end{framed}

\guidelinesubsubsection{Rationale}

LLM-based studies rest on artifacts that researchers design, author, or configure before any model is invoked: \emph{software layers} that pre-process data, prepare prompts, filter user requests, or post-process responses, and the \emph{context} those layers feed into each call~\cite{galster2026configuring}. Context includes prompt templates, context files, tool and skill schemas, and retrieval mechanisms that bring external data into model invocations.
For example, ChatGPT and GitHub Copilot use the same underlying models, but their outputs differ substantially because Copilot automatically adds project context. Researchers can also build tools using models directly via APIs.

Prompts are central to any LLM-based study~\cite{DBLP:conf/iclr/Sclar0TS24}.
A \emph{prompt} is a concrete input to an LLM that guides its output~\cite{DBLP:journals/corr/abs-2406-06608}.
Depending on the task, a prompt may include instructions (e.g., \enq{classify the following bug report}), task context, input data, and output format specifications (e.g., \enq{respond as JSON with fields `category' and `justification'}), with outputs ranging from unstructured text to structured formats such as JSON~\cite{promptingguide_elements}.
A \emph{prompt template} is a parameterized structure containing static elements (e.g., instructions, output format specifications) and placeholders for variable content (e.g., source code under analysis) that are filled in at runtime to construct concrete prompts~\cite{DBLP:journals/corr/abs-2406-06608}.
In automated studies, researchers typically design prompt templates from which individual prompts are then instantiated.
Prompts substantially influence a model's output, so how they are formatted and integrated into an LLM-based study is essential for transparency, verifiability, and reproducibility.
\citeauthor{DBLP:conf/iclr/Sclar0TS24} question the methodological validity of comparing models with \enq{an arbitrarily chosen, fixed prompt format}, because their research has shown that the performance of different prompt formats only weakly correlates between different models~\cite{DBLP:conf/iclr/Sclar0TS24}.

This guideline covers the architecture, prompts, and configuration used in an LLM-based study, complementing \modelversion (model-specific details) and \traces (runtime behavior).
It does not apply when LLMs are used solely for language polishing, paraphrasing, translation, tone or style adaptation, or layout edits (see \scope).

\guidelinesubsubsection{Recommendations}

Researchers \must clearly describe the tool architecture and what exactly the LLM (or ensemble of LLMs) contributes to the tool or method presented in a research paper, including any dependencies on proprietary tools that affect reproducibility.
Researchers \should justify substantive architectural choices where alternatives existed (e.g., why a particular agentic framework or tool catalog was selected).
Researchers \should describe how the models were hosted and accessed. For \emph{time-sensitive measurements}, the hosting choice (e.g., self-hosting on local hardware, an aggregator such as OpenRouter, or a vendor API such as the OpenAI API) can substantially affect results, so researchers \must clarify whether local infrastructure or cloud services were used, including the specific hardware for local hosting (e.g., GPU model and VRAM) or the service tier for cloud APIs (latency reporting is covered under \benchmarksmetrics).
Where legally possible (e.g., not restricted by industry-partner agreements), researchers \should release the source code of their implementation under an open-source license.

Reporting requirements scale with system complexity: minimal for standalone LLMs, detailed for pipelines, agentic systems, and any context files or tool schemas they depend on. In the topic-specific paragraphs that follow, the \paper \must contain a high-level description of any reported component, with full details provided as \supplementarymaterial.

\guidelineparagraph{Prompt Reporting}
Researchers \must report all prompts used in an empirical study, including instructions, task context, input data, and output indicators.
The complete set \must be made publicly available as \supplementarymaterial, with representative examples in the \paper itself.
When confidentiality (e.g., industry-partner agreements) prevents full publication, researchers \should publish summaries and representative examples instead.
For prompts that can be partially shared, researchers \must anonymize personal identifiers, replace proprietary code with placeholders, and clearly highlight modified sections.
When prompt templates are used, researchers \must report them alongside representative instances, specifying which parts are static and which are dynamically filled.
When prompts are generated dynamically, researchers \must document the code or rules that assemble each prompt from runtime inputs.
Researchers \should specify the exact formatting of prompts, including how code snippets were enclosed (e.g., triple backticks), whether whitespace was preserved, and how other artifacts such as error messages and stack traces were formatted.
For studies involving human participants who create or modify prompts, researchers \should describe how these prompts were collected and analyzed.

\guidelineparagraph{Prompt Development}
Researchers \should explain in the \paper how they developed the prompts and why they decided to follow certain prompting strategies.
If prompts from the early phases of a research project are unavailable, researchers \should at least summarize the prompt evolution.
Prompt development is often iterative, involving collaboration between human researchers and AI tools.
Researchers \should report any instances in which LLMs were used to suggest prompt refinements and how these suggestions were incorporated.
A prompt changelog can track prompt evolution, including revisions and reasons for changes (e.g., v1.0: initial prompt; v2.0: added few-shot examples).
Because prompt effectiveness varies between models and model versions, researchers \must make clear which prompts were used for which models in which versions and with which configuration.

\guidelineparagraph{Prompting Strategy and Input Handling}
Researchers \must specify whether zero-shot, one-shot, or few-shot prompting was used.
For few-shot prompts, researchers \must explain in the \paper how the examples were selected and \should include the concrete examples in the \supplementarymaterial.
If multiple versions of a prompt were tested, researchers \should describe how these variations were evaluated and how the final design was chosen.
When dealing with extensive or complex prompt context, researchers \should describe the strategies they used to handle input length constraints (e.g., truncating, summarizing, or splitting prompts into multiple parts).
Token optimization measures, such as simplifying code formatting or removing unnecessary comments, \should also be documented if applied.

\guidelineparagraph{Pipelines and Complex Systems}
If the LLM is used in a \emph{standalone setup}, with prompts sent directly to a model via an API and no pre-processing of inputs or post-processing of outputs, researchers \must state this explicitly.
If the LLM is part of a \emph{complex system} (e.g., with pre-processing or post-processing stages), researchers \must describe each component's role and the data flow between them.
For systems using retrieval-augmented generation (RAG) or related methods (e.g., rule-based retrieval, structured query generation, or hybrid approaches), researchers \must additionally describe how external data was retrieved, stored (e.g., in vector databases, knowledge graphs), and selected for inclusion in the model's context.
The data used for retrieval \should also be reported, including its preprocessing, versioning, and update frequency. If not confidential, an anonymized snapshot \should be made available as \supplementarymaterial.
For \emph{ensemble models}, in addition to following the \modelversion guideline for each model, researchers \should describe the architecture connecting them: the routing logic that determines which model handles which input, model interactions, and the output combination strategy (e.g., majority voting, weighted averaging, sequential processing).

\guidelineparagraph{Agentic Systems}
If the LLM is part of an \emph{agentic system} that autonomously plans or executes tasks, researchers \must additionally describe the agents' roles (e.g., planner, executor, coordinator), whether the system is single-agent or multi-agent, how the agents interact with external tools and users, and the reasoning framework used (e.g., chain-of-thought, self-reflection, multi-turn dialogue).
For agentic systems that use external tools (e.g., Claude Code and its \href{https://code.claude.com/docs/en/sub-agents}{subagents}), researchers \must distinguish three kinds of activity: (1) the model's reasoning, planning, and outputs; (2) tool calls (e.g., to APIs, databases, file systems, or \href{https://modelcontextprotocol.io/}{Model Context Protocol (MCP)} servers); (3) interactions with users, the environment, or other agents.
Reporting these separately lets readers understand whether a result came from the model, a tool, or their interaction. How to record the runtime traces of each is covered in \traces.

\guidelineparagraph{Context Files and Agent Configuration}
Researchers can tailor agentic tools through \emph{configuration mechanisms} such as context files, skills, subagents, hooks, settings, and rules~\cite{galster2026configuring}.
A \emph{configuration artifact} is a concrete instance of such a mechanism: a single file (e.g., a context file or a subagent file) or a directory bundling several files (e.g., a skill folder containing \texttt{SKILL.md} alongside scripts, references, and assets).
Since configuration mechanisms steer agent behavior in the same way as system prompts, researchers \must describe which configuration mechanisms were used and \should publish all configuration artifacts as \supplementarymaterial.
Configuration mechanisms and their artifacts \must be reported with the same level of detail as general prompts, including their development process and any iterations.
Because context files are version-controlled, their evolution across a study is recoverable from the project's Git history and from the runtime traces reported under \traces.
Where subagents are used, researchers \should also describe the delegation pattern, including which subagent handles which task and how control is returned to the orchestrator.

\guidelineparagraph{Tool Catalog and MCP Servers}
On each call, agentic systems expose a set of tools to the model along with their schemas (e.g., parameter names, types, and natural-language descriptions).
Tools differ in granularity. An editor agent might expose one tool per editing operation (e.g., \texttt{read\_file}, \texttt{apply\_edit}), while Claude Code defines a single \texttt{Bash} tool (\texttt{PowerShell} on Windows) through which the agent runs arbitrary shell commands such as \texttt{grep} or \texttt{ls}~\cite{anthropic_tools_reference}.
The catalog, the wording of descriptions, and the order in which tools are presented all influence which tool the model selects, so the exact serialized form matters for reproducibility.
The same applies to the list of MCP servers made available to the model.
Researchers \must summarize in the \paper which tools were exposed to the model, and \should include a complete list with names and purposes, tool schemas, and the names of any connected MCP servers as \supplementarymaterial.

\guidelinesubsubsection{Examples}

\citeauthor{DBLP:journals/tse/SchaferNET24} evaluated LLMs for automated unit test generation, providing a detailed description of the system architecture including code parsing, prompt formulation, LLM interaction, and test suite integration~\cite{DBLP:journals/tse/SchaferNET24}.
They also detailed the datasets used, including sources, selection criteria, and preprocessing steps.

A second example is \citeauthor{DBLP:conf/chi/YanHWH24}'s \emph{IVIE} tool~\cite{DBLP:conf/chi/YanHWH24}, which integrates LLMs into the VS Code interface.
The authors document the tool architecture, detailing the IDE integration, context extraction from code editors, and the formatting pipeline for LLM-generated explanations.

\citeauthor{DBLP:journals/pacmse/LiangBBDFF024}'s paper is a good example of comprehensive prompt reporting~\cite{DBLP:journals/pacmse/LiangBBDFF024}.
The authors make the exact prompts available in their \supplementarymaterial on Figshare, including details such as code blocks enclosed in triple backticks.
The \paper explains the rationale behind the prompt design and the data output format, and it includes an overview figure and two concrete examples, keeping the main text concise while remaining reproducible.

\guidelinesubsubsection{Benefits}

Documenting the tool architecture and hosting infrastructure of LLM-based systems strengthens reproducibility and transparency, enabling experiment replication, result validation, and cross-study comparison.
Prompt documentation provides similar benefits, letting other researchers replicate studies, refine prompts, and evaluate how content and formatting choices influence LLM behavior.
Reporting configuration mechanisms, the tool catalog, and MCP servers additionally lets other researchers reconstruct the configured context, often the dominant factor in agent behavior.

\guidelinesubsubsection{Challenges}

Documenting LLM-based architectures involves challenges such as proprietary APIs and dependencies that restrict disclosure, managing large-scale retrieval databases, and ensuring efficient query execution.
Researchers must also balance transparency with data privacy concerns, adapt to the evolving nature of LLM integrations, and handle the complexity of multi-agent interactions and decision-making logic, all of which can impact reproducibility and system clarity.
Prompts themselves are challenging to document because they often combine multiple components such as code, error messages, and explanatory text, and privacy or confidentiality concerns can hinder sharing.

Not all systems allow reporting of complete (system) prompts, context files, and tool schemas.
For commercial tools, researchers \must report all available information and acknowledge unknown aspects as limitations.
Disclosure practices vary across vendors: some publish their system prompts, others keep them proprietary.
Where prompts are published, they also change between releases~\cite{willison2026opus}, so researchers \should record the tool version and date of use even when the prompt content itself is unavailable.
Understanding suggestions of commercial tools such as \emph{GitHub Copilot} might require recreating the exact state of the codebase at the time the suggestion was made, which is a challenging context to report.
One solution is to use version control to capture the exact state of the codebase when a recommendation was made, keeping track of the files that were automatically added as context.
We also recommend exploring open-source tools such as \emph{OpenCode}~\cite{opencode}, which expose more of the configuration that controls agent behavior.

\guidelinesubsubsection{Study Types}

This guideline \must be followed for all studies that involve tools with system-level components beyond bare LLMs, from lightweight wrappers that pre-process user input or post-process model outputs, to systems employing retrieval-augmented methods or complex agentic architectures.
It also \must be followed by all studies that use concrete prompts or prompt templates.

For \newtools, this guideline is of primary importance: researchers \must describe the tool's full architecture, explain how prompts were generated and structured within the tool, report the context files, tool catalog, and skill definitions used, and document how the role of each LLM fits into the overall system behavior.
For \benchmarkingtasks, researchers \must describe the evaluation harness and infrastructure when it goes beyond bare model API calls (e.g., custom sandboxing, orchestration layers, or post-processing pipelines), and the harness \should be usable with open models; see~\cite{Anthropic2025evals} for a practitioner account of evaluation harness design for agentic systems.
Researchers using pre-defined prompts (e.g., \emph{HumanEval}, \emph{SWE-Bench}) \must specify the benchmark version and any modifications made to the prompts or evaluation setup.
If prompt tuning, RAG, or other methods were used to adapt prompts, researchers \must disclose and justify those changes, and \should make the relevant code publicly available.
For \llmusage, researchers \should describe the tool architecture of the studied tool to the extent that it is accessible, as architectural details may influence observed usage patterns.
For controlled experiments under \llmusage, exact prompts \must be reported for all conditions.
For \annotators, researchers \must document any predefined coding guides or instructions included in prompts, as these influence how the model labels artifacts.
For \judges, researchers \must report the evaluation criteria, scales, and examples embedded in the prompt to ensure consistent interpretation.
For \synthesis tasks (e.g., summarization, aggregation), researchers \must document the sequence of prompts used to generate and refine outputs, including follow-ups and clarification queries.
For \subjects (e.g., simulating human participants), researchers \must report any role-playing instructions, constraints, or personas used to guide LLM behavior.
For \annotators, \judges, \synthesis, and \subjects, if the research setup involves a custom pipeline (e.g., RAG for annotation or chained prompts for synthesis), the architecture \should also be reported.

\guidelinesubsubsection{Advice for Reviewers}

As with other guidelines, missing architectural or prompt information is typically a minor revision request unless so much is missing that methodological rigor cannot be assessed. Reviewers may ask authors to move key details from supplements into the paper body to ensure the main text is self-contained.

Regarding proprietary or confidential components, three principles apply: (1) empirically evaluating an opaque artifact is a valid scientific contribution; (2) the less available and inspectable the artifact, the weaker the contribution; (3) \textit{scientific instruments} including tools, metrics, scales, and experimental materials must be fully disclosed, even when the object under study cannot be.

A challenging aspect of prompt reporting is the description of prompt development, which is inherently iterative and creative. Reviewers should \textit{not} expect justification of each word choice or post-hoc rationalized accounts of prompt generation. Instead, reviewers should focus on whether (1) a new research team could \textit{use} (not reproduce) exactly the same prompts in exactly the same way, and (2) potential biases or validity issues are transparent.

\guidelinesubsubsection{See Also}
\begin{itemize}[label=$\rightarrow$]
    \item \seemodelversion: Every architecture runs on at least one specific model; name its version.
    \item \seetraces: Architecture and prompts are static; session traces show how they behave at runtime.
    \item \seehumanvalidation: Human validation often complements automated metrics for tool outputs.
    \item \seeopenllm: Open models let other researchers run the reported prompts and schemas on the exact same model weights.
    \item \seelimitationsmitigations: When a tool hides internal prompts or schemas, authors must report the gap as a limitation.
\end{itemize}
\space%

\subsection{Report Session Traces}
\label{sec:report-session-traces}

\vspace{0.5\baselineskip}
\begin{framed}
\summary: %
To address model non-determinism and ensure reproducibility, especially when targeting SaaS-based commercial tools, researchers \should include full interaction logs (prompts and responses) as \supplementarymaterial if privacy and confidentiality can be ensured. Traces containing sensitive information \must be anonymized. For agentic systems, interaction logs \should cover human-in-the-loop exchanges with the agent, including feedback and approval decisions. Researchers \should report the complete runtime trace as \supplementarymaterial, covering both external tool invocations (tool name, arguments, result, ordering) and which configured artifacts (skills, context files, subagents reported under \design) were activated, so that readers can attribute task outcomes to the model, the tools, or their interaction pattern. Where tool-native trace formats are used, the file format and tool version \must be described; an open format with a documented schema \should be preferred. Developed plans \should be reported as \supplementarymaterial if available. When full trace disclosure is not feasible, representative or anonymized examples \should be provided, and unobservable aspects of commercial-tool runs \must be acknowledged as reproducibility limitations.
\end{framed}

\guidelinesubsubsection{Rationale}

A \emph{session} is any bounded period of activity during which an LLM is invoked. Sessions cover one-shot prompts, batch runs, multi-turn conversations, and agentic runs that span many tool calls.
A \emph{session trace} records everything that crosses the LLM boundary or is produced around the model during that period: prompts received, responses returned, tool calls the model made together with their arguments and results, plans the model developed, and which of the statically configured artifacts (e.g., skills, context files, subagents, tools) were actually picked up at runtime.

Two kinds of session trace are important to capture.
An \emph{interaction log} captures what a human can observe at the LLM's interface: the prompts sent in and the responses returned, whether the prompts come from a human user or, in non-interactive runs, from a calling harness.
A \emph{runtime trace} records the LLM's internal activity: each call to an external tool (e.g., APIs, file systems, databases, MCP servers, subagents) or activation of a configured artifact (e.g., context files, skills), with the tool or artifact identified, the arguments passed, and the result returned. For agentic runs, both kinds matter, because neither tells the full story on its own.

Even with the exact same prompts, decoding strategies, and parameters, LLMs can behave non-deterministically.
Non-determinism can arise from probabilistic sampling and, even with greedy decoding (temperature = 0), from batching and floating-point arithmetic on GPUs~\cite{YuanNondeterminismLLMInference}, and from Mixture-of-Experts routing~\cite{Chann2023}.
Verifying conclusions drawn from such interactions therefore depends as much on runtime traces as on the system design itself.
This matters particularly for studies targeting commercial software-as-a-service (SaaS) solutions such as ChatGPT, and for agentic runs where behavior depends on how the model chose among many possible tool calls.

The rationale is similar to reporting interview transcripts in qualitative research.
Just as a human participant might give different answers to the same question asked two months apart, the responses from tools such as ChatGPT can also vary over time, and the trace of an agent's decisions on any given run is often not reproducible at a later date.
This guideline addresses runtime reporting and complements \design, which covers the static artifacts that determine the model's input.

\guidelinesubsubsection{Recommendations}

\guidelineparagraph{Interaction Logs}
Researchers \should report the full interaction logs (prompts sent to the LLM and responses returned) as part of their \supplementarymaterial.
For agentic systems, interaction logs cover the human-facing exchanges with the agent, including human-in-the-loop feedback, approval or rejection decisions, and iterative refinements.
These \should also be reported as \supplementarymaterial so that readers can reconstruct the sequence of exchanges and assess human oversight decisions.
For traces containing sensitive information, researchers \must anonymize personal identifiers, replace proprietary code with placeholders, and clearly highlight modified sections.

\guidelineparagraph{Runtime Traces}
When an LLM calls out to external tools (e.g., APIs, file systems, databases, MCP servers, subagents) or activates configured artifacts reported under \design (e.g., context files, skills, subagents), this runtime activity forms a \emph{runtime trace} distinct from the interaction log.
Researchers \should report the complete runtime trace as \supplementarymaterial, including for each entry the tool or artifact name, arguments (if any), result, and ordering relative to surrounding interaction-log entries.
This lets readers attribute task success to the model, the external tools, or their interaction pattern, and distinguish artifacts that were configured from those that actually influenced a given run.
Researchers \should use an open format with a documented schema. Emerging standards such as the OpenTelemetry GenAI semantic conventions~\cite{otel_genai_semconv} or OpenInference~\cite{openinference_spec} are preferred where they fit. Where tool-native formats are used (e.g., Claude Code's session transcripts, LangGraph's state logs), researchers \must describe the file format and report the tool version.

\guidelineparagraph{Agentic Plans}
For agentic systems that autonomously plan and execute tasks, researchers \should report any plans the system exposes as \supplementarymaterial.
In Claude Code, for example, a plan is a short Markdown document the user can open and edit during a session, listing the proposed steps and the files or commands the agent intends to touch.
Other frameworks such as LangGraph keep plans inside the agent's internal execution state.
All reported traces \must be made publicly available as \supplementarymaterial, subject to privacy and confidentiality constraints. When full trace logging is not feasible, researchers \should provide representative examples or anonymized traces.

\guidelinesubsubsection{Examples}

An example of reporting full interaction logs is the study by \citeauthor{ronanki2023investigating}, for which the authors reported the full answers of ChatGPT and uploaded them to \href{https://zenodo.org/records/8124936}{Zenodo}~\cite{ronanki2023investigating}.
For agentic systems, \citeauthor{DBLP:conf/kbse/BouzeniaP25} unified the runtime trajectories of three SE agents (\emph{RepairAgent}, \emph{AutoCodeRover}, \emph{OpenHands}) into a custom thought-action-result format and released the resulting 120 trajectories with 2,822 LLM interactions as a public dataset~\cite{DBLP:conf/kbse/BouzeniaP25}.

\guidelinesubsubsection{Benefits}

Unlike human participant conversations, which often cannot be reported because of confidentiality, LLM interaction logs can be shared.
This enables reproduction studies, tracking of response changes over time or across model versions, and secondary research on LLM consistency for specific SE tasks.

For agentic systems, reporting runtime traces alongside interaction logs lets readers follow the model's reasoning, the tool calls it made, and the order of those calls.
Runtime traces complement the static configuration reported under \design by showing which of the configured artifacts were activated on a given run.

\guidelinesubsubsection{Challenges}

Not all systems allow reporting of complete interaction logs with ease, and this hinders transparency and verifiability.
For commercial tools, researchers \must report all available information and acknowledge unknown aspects as limitations.
Tool-call traces for commercial SaaS agents are often opaque: the user sees the final response but not the sequence of internal tool calls.
When this is the case, authors \should document what was and was not observable.
For agents running locally or via open-source tools, these traces are usually more accessible and \should be reported whenever available.
Agent frameworks differ in whether and how they log agent-to-agent communication, so reporting practices vary across studies.

\guidelinesubsubsection{Study Types}

Runtime trace reporting requirements depend heavily on the study type and on the accessibility of the underlying system.
The general expectation is that interaction logs are reported as \supplementarymaterial whenever feasible; runtime traces are additionally expected when agentic execution or evaluation harnesses go beyond direct API calls.

For \llmusage, especially observational studies targeting commercial tools, researchers \must report the full interaction logs except when transcripts might identify anonymous participants or reveal personal or confidential information.
If complete interaction logs cannot be shared (e.g., because they contain confidential information), the prompts and responses \must at least be summarized and described in the \paper.
For \newtools that use agentic execution, researchers \should report runtime traces for the runs used to evaluate the tool.
For \benchmarkingtasks that use agent-based harnesses, researchers \should report runtime traces for representative runs, letting readers understand how task success depends on the agent's decision sequence rather than the model's raw output alone.
For \annotators, \judges, \synthesis, and \subjects, when the research setup involves multi-turn interaction or agentic orchestration, researchers \should report the corresponding interaction logs and, where applicable, runtime traces.

\guidelinesubsubsection{Advice for Reviewers}

As with other guidelines, missing trace information is typically a minor revision request unless so much is missing that methodological rigor cannot be assessed.
Reviewers should recognize that complete trace reporting is easier for local and open-source setups than for commercial SaaS agents.
When reviewing commercial-tool studies, reviewers should focus on whether authors have reported everything the system exposed and have been explicit about what remains unobservable.

\guidelinesubsubsection{See Also}
\begin{itemize}[label=$\rightarrow$]
    \item \seedesign: Session traces show runtime behavior; system and prompt design covers the static artifacts that produce it.
    \item \seemodelversion: A trace cannot be reproduced or compared across studies without the model identity that produced it.
    \item \seehumanvalidation: Agentic traces produce outputs that often warrant human validation.
    \item \seeopenllm: Open models let other researchers reproduce a reported trace on the exact same model weights.
    \item \seelimitationsmitigations: When a tool hides parts of the trace, authors must report the gap as a limitation.
\end{itemize}
\space%

\subsection{Use Suitable Baselines, Benchmarks, and Metrics}
\label{sec:use-suitable-baselines-benchmarks-and-metrics}

\vspace{0.5\baselineskip}
\begin{framed}
\summary: %
Researchers \must justify all benchmark and metric choices in the \paper, \must discuss their reliability and validity (especially construct validity), and \should summarize benchmark structure, task types, and limitations. They \should operationally define the phenomenon being measured, justify the sampling strategy used to select problems for inclusion in the benchmark, isolate the target capability from confounders where possible, and perform an error analysis. When creating or releasing a benchmark, data sources and collection dates \must be disclosed for each release. Where possible, traditional (non-LLM) baselines \should be used for comparison. Researchers \must explain in the \paper why the selected metrics are suitable for the specific study; prior adoption in related work alone does not constitute sufficient justification. They \should report established metrics to make study results comparable, but can report additional metrics that they consider appropriate. Due to the inherent non-determinism of LLMs, experiments \should be repeated; the result distribution \should then be reported using descriptive statistics. If comparing models or tools, researchers \should use appropriate inferential statistics rather than relying solely on summary statistics. Researchers \should justify the number of experiment repetitions, for example through a power analysis or by monitoring convergence of descriptive statistics. Latency \must be reported when it can affect study outcomes (e.g., interactive user studies, latency comparisons).
\end{framed}

\guidelinesubsubsection{Rationale}

Meaningful evaluation requires well-understood, valid measurement instruments.
Without justified benchmarks and metrics, claims about LLM performance lack the rigor needed for scientific comparison and cumulative progress.
A \textit{benchmark} is a \enq{standard tool for the competitive evaluation and comparison of competing systems or components according to specific characteristics, such as performance, dependability, or security}~\cite{DBLP:conf/wosp/KistowskiAHLHC15}. A \textit{metric} \enq{is a method, algorithm, or procedure for assigning one or more numbers to a phenomenon}~\cite{DBLP:books/sp/24/RalphKAA24}. A baseline is a reference point, enabling comparison of LLMs against traditional algorithms with lower computational costs.

\guidelinesubsubsection{Recommendations}

Benchmark and metric selection requires understanding the benchmark tasks, what exactly is being measured, and how it relates to the (often latent) variables researchers actually care about. When one or more metrics or benchmarks are used, researchers \must briefly justify in the \paper why each selected benchmark and metric is suitable for the given task or study, and \must discuss the reliability and validity---especially construct validity---of those choices (see \limitationsmitigations). Beyond these requirements, researchers \should:
\begin{itemize}
    \item provide an operational definition of the phenomenon the benchmark is intended to measure (e.g., functional correctness, code maintainability, vulnerability detection), including its scope and any sub-components~\cite{bean2025measuring};
    \item summarize the structure and tasks of the selected benchmark(s), including the programming language(s) and descriptive statistics such as the number of contained tasks and test cases;
    \item describe and justify the sampling strategy used to select problems for inclusion in the benchmark (e.g., function-completion problems in \emph{HumanEval}, GitHub issues in \emph{SWE-bench}, vulnerable functions in \emph{DiverseVul}); if non-probability sampling (e.g., convenience) is used, discuss its implications for the generalizability of conclusions~\cite{DBLP:journals/ese/BaltesR22, bean2025measuring};
    \item discuss the limitations of the selected benchmark(s) (e.g., widely used benchmarks such as \emph{HumanEval} \cite{DBLP:journals/corr/abs-2107-03374} and \emph{MBPP} \cite{DBLP:journals/corr/abs-2108-07732} only test short Python functions, which is not representative of the full breadth of SE work~\cite{Chandra2025benchmarks});
    \item include an example of a task and the corresponding test case(s) to illustrate the structure of the benchmark.
\end{itemize}

If multiple benchmarks exist for the same task, researchers \should compare both performance and design choices (e.g., which tasks are included, how outputs are scored, what aspect of the phenomenon is covered) across benchmarks~\cite{bean2025measuring}.
When selecting only a subset of all available benchmarks, researchers \should use the most specific benchmarks given the context.
When adapting an existing benchmark, researchers \should document what was changed and why, and \should report performance on both the original and the adapted version where feasible~\cite{bean2025measuring}.

Benchmark scores often confound the target capability with unrelated capabilities the task happens to require. For code translation benchmarks, prompt formatting alone can shift performance by up to 40\%~\cite{DBLP:journals/corr/abs-2411-10541, cao2025should}, conflating prompt format sensitivity with translation capability. More broadly, output format compliance (e.g., specific JSON schemas or unit-test conventions) and general instruction-following capability are routinely bundled into the aggregate benchmark score~\cite{bean2025measuring}. Researchers \should identify which capabilities a benchmark conflates, isolate the target capability where possible (e.g., by reporting per-subtask breakdowns or by adopting benchmarks designed to test the target capability in isolation), and acknowledge remaining confounders as threats to construct validity.

After running a benchmark, researchers \should perform an error analysis by categorizing the failures observed and reporting the relative frequency of each category. If failures concentrate on inputs that demand capabilities other than the target capability (e.g., reading across many files rather than fixing the bug the benchmark targets), this is a construct-validity threat and \should be reported alongside the primary scores~\cite{bean2025measuring}.

In addition to disclosing data sources and collection dates (see \emph{Challenges} below), researchers creating or releasing a new benchmark \should adopt concrete contamination-prevention mechanisms: maintaining a held-out subset of items for ongoing, uncontaminated evaluation; embedding canary strings, that is, unique markers that downstream tools can later search for in model outputs to detect inclusion in training data; and investigating whether the benchmark's source materials may already appear in common LLM training corpora~\cite{bean2025measuring, cao2025should}. Researchers using existing benchmarks \should additionally discuss contamination as a study limitation (see \limitationsmitigations).

Furthermore, researchers \should check whether a less resource-intensive approach (e.g., for static analysis tasks or program repair) can serve as a baseline. If so, the LLM or LLM-based tool \should be compared with such baselines using suitable metrics. Even if LLM-based tools outperform baselines, researchers \should discuss whether the resources consumed justify the (often marginal) improvements~\cite{DBLP:journals/cacm/Menzies25}.

To compare traditional and LLM-based approaches or different LLM-based tools, researchers \should report established metrics whenever possible, as this allows secondary research.
They can report additional metrics that they consider appropriate.
We briefly discuss common metrics in the \emph{Examples} subsection below.
As mentioned, researchers \must motivate why they chose a certain metric or variant thereof for their particular study.
Prior adoption alone does not constitute sufficient justification; researchers \mustnot justify metric choices solely by citing their use in prior work.
Latency \must be reported when it can affect study outcomes.

Due to LLM non-determinism, researchers \should repeat experiments and report descriptive statistics of model or tool performance (e.g., arithmetic mean, median, confidence intervals, standard deviations)~\cite{DBLP:conf/nips/AgarwalSCCB21, DBLP:journals/corr/abs-2602-07150}. If comparing models or tools, researchers \should use appropriate inferential statistics with effect sizes rather than relying solely on differences in means or other summary statistics. Suitable choices include hypothesis tests such as the Mann-Whitney U test, McNemar's test for binary outcomes~\cite{DBLP:conf/iclr/KublerBKZYKK26}, and bootstrap-based comparisons. For choosing among these and related tests, \citeauthor{DBLP:conf/acl/DrorBSR18} provide a decision tree based on the distributional assumptions of the test statistic and the size of the test set~\cite{DBLP:conf/acl/DrorBSR18}. When scores vary across raters or across runs of the same scorer (e.g., multiple human raters or repeated runs of an LLM judge), researchers \should report the distribution of ratings per item rather than only aggregated point estimates or exact-match agreement, since aggregation can mask systematic disagreement~\cite{bean2025measuring}.

The number of required repetitions depends on factors such as the study type, the variability of the task, and the desired precision of estimates. As with sample sizes for human validation (see \humanvalidation), researchers \should justify their chosen number of repetitions, for example, through a power analysis~\cite{Cohen1992, DBLP:journals/corr/abs-2602-07150} or by monitoring the convergence of descriptive statistics across incremental runs~\cite{DBLP:journals/corr/abs-2410-03492}. A pilot study can help estimate the expected variability and inform this decision.

From a measurement perspective, researchers \should reflect on the theories, values, and measurement models on which the benchmarks and metrics they have selected for their study are based.
For example, labeling phenomena as ``bugs'' in a large open dataset reflects a certain theory of what constitutes a bug, as well as the values and perspectives of the people who labeled the dataset. Reflecting on the context in which these labels were assigned and discussing whether and how the labels generalize to a new study context is crucial.

Researchers building or releasing new SE benchmarks \should consult operational checklists. \citeauthor{cao2025should} provide \emph{HOW2BENCH}, a 55-item checklist covering benchmark design, construction, evaluation, analysis, and release~\cite{cao2025should}. \citeauthor{bean2025measuring} provide a complementary domain-agnostic checklist organized around the construct-validity recommendations summarized above~\cite{bean2025measuring}.

\guidelinesubsubsection{Examples}

\guidelineparagraph{Common Metrics}

Two common metrics used for generation tasks are \emph{BLEU-N} and \emph{pass@k}.
\emph{BLEU-N}~\cite{DBLP:conf/acl/PapineniRWZ02} was originally developed for evaluating machine translation quality by measuring modified n-gram precision (with a brevity penalty) between a candidate and reference text, ranging from 0 (dissimilar) to 1 (similar). It has been widely adopted in SE for code generation tasks, though its validity in this context is debatable. An n-gram overlap does not capture functional correctness, and syntactically different code can be semantically equivalent (see \emph{Challenges} below).
Code-specific variations attempt to address these limitations. \emph{CodeBLEU}~\cite{DBLP:journals/corr/abs-2009-10297} augments n-gram overlap with syntactic (AST) and data-flow matching, whereas \emph{CrystalBLEU}~\cite{DBLP:conf/kbse/EghbaliP22} ignores n-grams that recur across unrelated programs, such as boilerplate syntax and common API calls, because they inflate overlap scores without indicating genuine similarity.

The metric \emph{pass@k} reports the probability that at least one of $k$ generated solutions for a task passes the task's tests.
In their pseudocode-to-code study, \citeauthor{DBLP:conf/nips/KulalPC0PAL19} counted a test example as solved if best-first search over one candidate code line per pseudocode line produced an accepted program within $B$ trials.
Each trial compiled one candidate program and, if compilation succeeded, ran public tests~\cite{DBLP:conf/nips/KulalPC0PAL19}.
\citeauthor{DBLP:journals/corr/abs-2107-03374} defined \emph{pass@k} for code generation and estimated it from $n \ge k$ samples, treating a sample as correct when it passes the task's unit tests~\cite{DBLP:journals/corr/abs-2107-03374}.

For a single task, the estimator for \emph{pass@k} is:

\[
\text{pass@}k = 1 - \frac{\binom{n-c}{k}}{\binom{n}{k}},
\]
where:

\begin{itemize}
  \item $n$ is the total number of samples generated per task (with $n \ge k$),
  \item $c$ is the number of correct samples among the $n$, and
  \item $k$ is the number of attempts considered, drawn from the $n$ generated samples without replacement.
\end{itemize}

A benchmark reports the mean of the task-level \emph{pass@k} estimates.
The choice of $k$ depends on the downstream task. The metric \emph{pass@1} is critical for single-suggestion scenarios such as code completion, while higher $k$ values (e.g., 2, 5, 10) assess multi-attempt capability and are commonly reported in technical reports of code LLMs (e.g., Code Llama~\cite{DBLP:journals/corr/abs-2308-12950}, DeepSeek-Coder~\cite{DBLP:journals/corr/abs-2401-14196}, Qwen2.5-Coder~\cite{DBLP:journals/corr/abs-2409-12186}, StarCoder~\cite{DBLP:journals/tmlr/LiAZMKMMALCLZZW23}).
However, \emph{pass@k} is not a universal metric suitable for all generation tasks.
It requires a binary notion of correctness, making the metric appropriate for code synthesis evaluated via unit tests, but unsuitable for open-ended generation tasks such as comment generation, where multiple valid outputs exist.

A complementary perspective from industry practice is \emph{pass\textsuperscript{k}} (also written \emph{pass\^{}k}), which measures the probability that \emph{all} $k$ trials succeed rather than \emph{at least one}~\cite{Anthropic2025evals}. While \emph{pass@k} increases with $k$, \emph{pass\textsuperscript{k}} decreases, making it useful for assessing reliability in deployment scenarios where consistent success matters (e.g., customer-facing agents).

If a study evaluates an LLM-based tool for supporting humans, a relevant metric is the acceptance rate, meaning the ratio of all accepted artifacts (e.g., test cases, code snippets) in relation to all artifacts that were generated and presented to the user.
Another way of evaluating LLM-based tools is calculating inter-model agreement, which reveals how much a tool's performance depends on specific models and versions.
Metrics used to measure inter-model agreements include general agreement (percentage), \emph{Cohen's} $\kappa$, and \emph{Krippendorff's} $\alpha$ (see \humanvalidation for recommended thresholds and best practices for measuring agreement).

Common problem types for LLM-based studies are classification, recommendation, and generation, each requiring different metrics~\cite{DBLP:journals/tosem/HouZLYWLLLGW24}. \citeauthor{DBLP:journals/corr/abs-2505-08903} categorized 191 LLM benchmarks by SE task, providing a valuable reference~\cite{DBLP:journals/corr/abs-2505-08903}. For an overview of code foundation models, agents, and their evaluation, see~\citeauthor{yang2025code}~\cite{yang2025code}. From a practitioner perspective, \citeauthor{Anthropic2025evals} categorize evaluation approaches for LLM-based agents into code-based graders (e.g., test suite execution, static analysis), model-based graders (e.g., rubric-scored LLM judgments), and human graders~\cite{Anthropic2025evals}. This taxonomy may help researchers systematically design evaluation strategies for agent-based tools. Common metrics include \emph{BLEU}, \emph{pass@k}, \emph{Accuracy@k}, and \emph{Exact Match} for generation; \emph{Mean Reciprocal Rank} for recommendation; and \emph{Precision}, \emph{Recall}, \emph{F1-score}, and \emph{Accuracy} for classification.

\guidelineparagraph{Benchmark Examples}

Benchmarks used for code generation include \emph{HumanEval} (available on \href{https://github.com/openai/human-eval}{GitHub}) \cite{DBLP:journals/corr/abs-2107-03374}, \emph{MBPP} (available on \href{https://huggingface.co/datasets/google-research-datasets/mbpp}{Hugging Face}) \cite{DBLP:journals/corr/abs-2108-07732},
\emph{ClassEval} (available on \href{https://github.com/FudanSELab/ClassEval}{GitHub}) \cite{DBLP:conf/icse/Du0WWL0FS0L24}, \emph{LiveCodeBench} (available on \href{https://github.com/LiveCodeBench/LiveCodeBench}{GitHub}) \cite{DBLP:journals/corr/abs-2403-07974}, and \emph{SWE-bench} (available on \href{https://github.com/swe-bench/SWE-bench}{GitHub}) \cite{DBLP:conf/iclr/JimenezYWYPPN24}.
An example of a code translation benchmark is \emph{TransCoder}~\cite{DBLP:conf/nips/RoziereLCL20} (available on \href{https://github.com/facebookresearch/CodeGen}{GitHub}).
\citeauthor{DBLP:journals/corr/abs-2601-11895}'s \emph{DevBench} (available on \href{https://github.com/microsoft/devbench}{GitHub}) synthesized 1,800 code completion instances from developer telemetry, rather than collecting them from public sources~\cite{DBLP:journals/corr/abs-2601-11895}.
For evaluating LLMs as agents, \emph{AgentBench} (available on \href{https://github.com/THUDM/AgentBench}{GitHub})~\cite{DBLP:conf/iclr/0036YZXLL0DMYZ024} evaluates LLM agents across eight environments, including an operating system shell, a database, and a web browser.

Most benchmarks focus on generation tasks, but benchmarks for classification and recommendation also exist.
For classification, \emph{DiverseVul} (available on \href{https://github.com/wagner-group/diversevul}{GitHub})~\cite{DBLP:conf/raid/0001DACW23} provides vulnerable and non-vulnerable functions evaluated using standard classification metrics.
For recommendation, \emph{CodeSearchNet} (available on \href{https://github.com/github/CodeSearchNet}{GitHub})~\cite{DBLP:journals/corr/abs-1909-09436} contains code-documentation pairs evaluated using \emph{Mean Reciprocal Rank}.

\guidelinesubsubsection{Benefits}

Established benchmarks and metrics let researchers assess new systems against shared reference points rather than against hundreds of competing systems, showing which approaches work best on those benchmarks. Researchers can also track progress by iteratively improving a new LLM-based tool and re-testing it against the same benchmarks after substantial changes. For practitioners, leaderboards support selecting models for downstream tasks, and baselines clarify how much LLMs improve over non-LLM alternatives.

\guidelinesubsubsection{Challenges}

Computer scientists commonly use simple metrics (e.g., lines of code, CPU time) as proxies for complex, multidimensional latent variables (e.g., system size, environmental sustainability), without empirically validating that the metrics capture the intended construct~\cite{DBLP:conf/ease/RalphT18}. Unlike fields such as psychology where measurement theory has a longer tradition~\cite{borsboom2005measuring}, many AI metrics and benchmarks lack both theoretical underpinnings and empirical validation of their construct, measurement, and ecological validity.

These validity issues are widespread, not isolated. In a survey of 572 code benchmarks released between 2014 and 2025, \citeauthor{cao2025should} found that 84.2\% did not consider test suite coverage when constructing test cases, 64.0\% reported single-pass evaluations without controlling for randomness, and 82.5\% did not address data contamination~\cite{cao2025should}. \citeauthor{bean2025measuring} reported comparable patterns in a parallel review of 445 LLM benchmarks from ML and NLP venues~\cite{bean2025measuring}. As a result, model performance on these benchmarks can reflect benchmark artifacts rather than the capability the benchmark claims to test.

For example, \emph{pass@k} is intended to measure a model's \textit{functional correctness}, that is, its capability to generate code that produces the expected output for a given specification. However, functional correctness is only one dimension of code quality. \emph{pass@k} does not capture maintainability, readability, security, or efficiency of the generated code, all of which are critical for downstream use. Furthermore, ``correctness'' is defined entirely by test suites, whose coverage is itself unvalidated. A solution that passes all provided tests may still be incorrect for untested inputs. Whether a test suite adequately operationalizes correctness for a given task is a subjective judgment that is rarely examined.

Similarly, \emph{HumanEval} is intended to measure a model's capability to \textit{synthesize short Python functions from docstring specifications}. This operationalizes a narrow slice of ``code generation capability'': it covers neither multi-file tasks, nor debugging, nor the use of existing codebases. These tasks dominate real-world software engineering~\cite{Chandra2025benchmarks}. Notably, neither \emph{pass@k} nor \emph{HumanEval} has undergone rigorous empirical validation of its construct validity. Their widespread adoption rests on face validity and convenience rather than on evidence that they reliably measure what researchers intend them to measure~\cite{cao2025should}. \emph{HumanEval} also contains implementation, documentation, and test case bugs in its original release, which directly affect score interpretation~\cite{DBLP:conf/nips/LiuXW023}; comparable issues have been documented in other widely-used code benchmarks~\cite{cao2025should}.

Benchmarks and metrics also have generalizability problems. Prominent LLM benchmarks such as \emph{HumanEval} and \emph{MBPP} use Python, so researchers can optimize for Python's idiosyncrasies, producing gains that do not generalize to other languages. Similarly, the metric \emph{BLEU-N} is a syntactic metric, so code can score highly without being executable. The metric \emph{Exact Match}, meanwhile, does not account for functional equivalence of syntactically different code. Both \emph{BLEU-N} and \emph{Exact Match} are influenced by code formatting, which confounds their intended use.

Execution-based metrics such as \emph{pass@k} directly evaluate correctness by running test cases, but they require an execution environment.
When metric values look surprising, examine the specific test cases driving them: common causes include outliers, bugs in test suites or scoring code, and items that exercise capabilities other than the one being measured.

Finally, benchmark data contamination, where the benchmark itself is part of the training data, may lead to artificially high performance if the model remembers the solution from the training data rather than deriving it from the input~\cite{DBLP:journals/corr/abs-2406-04244} (see \limitationsmitigations). Therefore, for proprietary LLMs that do not release their training data, researchers \should consider using human validation, curating new data, or refactoring existing data~\cite{DBLP:journals/corr/abs-2406-04244}. 

To mitigate contamination, researchers can create new benchmark datasets by collecting data after a specified cutoff date; researchers \must disclose data sources and collection dates for each release. However, this temporal approach requires continuous updates as new models may include the benchmark in future training data. Alternatively, keeping the benchmark private prevents inclusion in training sets, but requires trust in the benchmark creator and a system to execute it without leaking data. A third strategy is to synthesize benchmark instances rather than draw them from public sources, which avoids training-data contamination at the cost of requiring an instance generator and a validation pipeline. Researchers can also evaluate how contaminated their benchmark is~\cite{DBLP:journals/corr/abs-2502-00678}.

\guidelinesubsubsection{Study Types}

This guideline \must be followed for all study types that automatically evaluate the performance of LLMs or LLM-based tools.
The design of a benchmark and the selection of appropriate metrics are highly dependent on the specific study type and research goal.
Recommending specific metrics for specific study types is beyond the scope of these guidelines, but \citeauthor{DBLP:journals/corr/abs-2505-08903} provide a good overview of existing metrics for evaluating LLMs~\cite{DBLP:journals/corr/abs-2505-08903}.

For \benchmarkingtasks, this guideline is of primary importance: researchers \must use established benchmarks or rigorously justify the creation of new ones, \must report standard metrics to enable cross-study comparison, and \should report an error analysis alongside the primary scores so that readers can assess whether reported gains reflect the target capability or other capabilities the task happens to require. When researchers compare LLMs or tools on latency, they \must report the infrastructure used to produce the measurements (\design).
For \annotators, the research goal might be to assess which model comes close to a ground truth dataset created by human annotators.
Especially for open annotation tasks, selecting suitable metrics to compare LLM-generated and human-generated labels is important.
In general, annotation tasks can vary widely.
Are multiple labels allowed for the same sequence? Are the available labels predefined, or should the LLM generate a set of labels independently?
Due to this task dependence, researchers \must justify their metric choice, explaining what aspects of the task it captures together with known limitations.
For \newtools, researchers \should benchmark the tool against suitable baselines using appropriate metrics that capture the tool's intended contribution.
For \judges, researchers \should report inter-rater agreement metrics and validity measures to demonstrate the reliability and quality of LLM judgments.
For \synthesis, researchers \should specify metrics for comparing synthesized outputs (e.g., coverage, faithfulness) and justify their appropriateness for the synthesis task.
For \llmusage, researchers \should justify the measurement instruments and metrics used for studying LLM usage patterns, including any survey scales or behavioral measures. In interactive setups where response times can influence participant behavior, researchers \must report observed latency.
For \subjects, researchers \should compare simulated and real human responses using appropriate metrics to assess simulation fidelity.
If researchers assess a well-established task such as code generation, they \should report standard metrics such as \emph{pass@k} and compare the performance between models.
If non-standard metrics are used, researchers \must state their reasoning.

\guidelinesubsubsection{Advice for Reviewers}

Reviewers should expect manuscripts to:
(1) clearly identify the constructs or variables the study aims to measure (e.g., LLM performance, quality of generated code), including independent, dependent, and control variables;
(2) present their measurement model, i.e., which metrics, benchmarks, or baselines are used and how they relate to the target constructs;
(3) justify the selection of metrics, benchmarks, and baselines;
(4) discuss \textit{in detail} the assumptions, reliability, and validity (\textit{especially construct validity}) of each benchmark and metric;
(5) articulate any limitations regarding construct and measurement validity.

As with other guidelines, missing information about baselines or metrics is typically a revision request. However, vague descriptions that conflate broad concerns (e.g., effectiveness, quality) with specific counting methods should be questioned. Ubiquity of a benchmark or metric does not imply validity or appropriateness for a given context. Manuscripts should convey a solid understanding of construct and measurement validity by explaining and justifying their measurement models.

\guidelinesubsubsection{See Also}
\begin{itemize}[label=$\rightarrow$]
    \item \seeopenllm: Using an open LLM as a baseline supports reproducible benchmark comparisons across studies.
    \item \seehumanvalidation: Human evaluation captures aspects of LLM outputs that automated metrics cannot measure.
    \item \seelimitationsmitigations: Every benchmark and metric choice carries construct-validity threats that authors must discuss.
\end{itemize}
\space%

\subsection{Use an Open LLM as a Baseline}
\label{sec:use-an-open-llm-as-a-baseline}

\vspace{0.5\baselineskip}
\begin{framed}
\summary: %
When researchers can control which LLM a study uses, they \should include an open-weight LLM, ideally an open-source LLM, as a baseline and report inter-model agreement. Researchers \should ensure the open-weight baseline is independently reproducible from their \supplementarymaterial.
\end{framed}

\guidelinesubsubsection{Rationale}

Reproducibility depends on access to the model under study.
When research relies exclusively on proprietary models, other researchers cannot independently verify or build upon the findings.
Including an open LLM as a baseline ensures that at least part of the study can be fully replicated.

\guidelinesubsubsection{Recommendations}

The term ``open'' can have different meanings in the context of LLMs.
\citeauthor{widder2024open} discuss three types of openness (i.e., transparency, reusability, and extensibility) and what openness can and cannot provide~\cite{widder2024open}.
We distinguish open-weight from open-source LLMs.
An \emph{open-weight} LLM has publicly released trained weights, so others can fine-tune, adapt, or deploy the model~\cite{OSIOpenWeights}.
An \emph{open-source} LLM, as defined by the \emph{Open Source Initiative} (OSI)~\cite{OSIAI2024}, also releases the complete source code used to train and run the model and sufficiently detailed information about the training data to build a substantially equivalent model.
These components are available under terms that grant the freedoms to use, study, modify, and share the model.
In our guidelines, an \emph{open LLM} is any LLM that is at least open-weight.
Open-source LLMs are preferable, because researchers can check their training data for benchmark contamination (see \benchmarksmetrics).

Empirical studies using LLMs in SE, especially those that target commercial tools or models, \should incorporate an open LLM as a baseline and report established metrics for inter-model agreement (see \benchmarksmetrics).
We acknowledge that including an open LLM baseline might not always be possible, for example, if the study involves human participants, and letting them work on the tasks using two different models might not be feasible.
Using an open model as a baseline is also not necessary if the use of the LLM is tangential to the study goal.

Open models allow other researchers to verify research results and build upon them, even without access to commercial models.
A comparison of commercial and open models also allows researchers to contextualize model performance.
Researchers \should ensure the open-LLM baseline is independently reproducible from their \supplementarymaterial.

Open LLMs are available from hubs such as \href{https://huggingface.co/}{\emph{Hugging Face}}. They can be self-hosted with frameworks such as \href{https://ollama.com/}{\emph{Ollama}} or \href{https://lmstudio.ai/}{\emph{LM Studio}}, accessed through cloud services such as \href{https://together.ai/}{\emph{Together AI}}, AWS, Azure, and Google Cloud, or routed through aggregators such as \href{https://openrouter.ai/}{\emph{OpenRouter}} that expose many providers behind a single API. For agentic setups, open-source tools such as \href{https://www.continue.dev/}{\emph{Continue}}, \href{https://cline.bot/}{\emph{Cline}}, and \href{https://opencode.ai/}{\emph{opencode}}~\cite{opencode} publish their full agent code, system prompts, and tool catalog. By contrast, vendor-hosted services such as \emph{GitHub Copilot} and \emph{Claude Code} expose only parts of their tooling (e.g., editor extensions, hook examples) and keep their deployed agent loops, system prompts, and tool catalogs proprietary.

\guidelinesubsubsection{Examples}

An increasing number of studies have adopted open LLMs as baseline models.
For example, \citeauthor{DBLP:journals/corr/abs-2406-09834} evaluated seven advanced LLMs, six of which were open-weight, testing 145 API mappings drawn from eight popular Python libraries across 28,125 completion prompts aimed at detecting deprecated API usage in code completion~\cite{DBLP:journals/corr/abs-2406-09834}.
\citeauthor{DBLP:journals/corr/abs-2408-04430} compared four LLMs on a cross-language code clone detection task~\cite{DBLP:journals/corr/abs-2408-04430}.
Three evaluated models were open-weight.
\citeauthor{DBLP:conf/eicc/GoncalvesSC0MPS25} fine-tuned the open LLM \emph{LLaMA} 3.2 on a refined version of the \emph{DiverseVul} dataset to benchmark vulnerability detection performance~\cite{DBLP:conf/eicc/GoncalvesSC0MPS25}.
\citeauthor{DBLP:journals/corr/abs-2601-11895} evaluated nine code completion models on the \emph{DevBench} benchmark and included three open-weight models (\emph{DeepSeek-V3}, \emph{DeepSeek-V3.1}, and \emph{Ministral-3B}) alongside commercial frontier models, releasing benchmark, evaluation scripts, and per-model raw completions under an MIT license~\cite{DBLP:journals/corr/abs-2601-11895}.
\emph{CodeBERT}, a bimodal transformer pre-trained by Microsoft Research, is published with model weights, source code, and data processing scripts on GitHub~\cite{codebert}. It has been used as an open baseline across diverse SE tasks, including exploit code generation~\cite{DBLP:journals/jss/YangZCZHC23}, vulnerability detection~\cite{DBLP:conf/gaiis/XiaSD24}, code clone detection~\cite{DBLP:conf/kbse/SonnekalbGBM22}, and programming assistance for exception handling~\cite{DBLP:conf/icse/CaiYMMN24}.

\guidelinesubsubsection{Benefits}

An open LLM baseline improves reproducibility by exposing model architectures, weights, and ideally training data, enabling independent verification of results.
Such baselines also let researchers compare novel methods against a stable reference point, since proprietary models can silently change between tests.
They also allow inspection of training data (when released) and model behavior, helping identify biases and limitations.
Unlike proprietary alternatives, which can be withdrawn or silently updated, open LLMs remain available for future studies.
They typically avoid the per-use API fees that can constrain research groups with limited budgets.

\guidelinesubsubsection{Challenges}

Open LLMs face several challenges:
\begin{itemize}
    \item \emph{Definitional inconsistency.} Many models that claim to be open or open source, such as Meta's \emph{Llama} and Google's \emph{Gemma}, release only trained weights without training data or methodological details~\cite{Gibney2024}, which is why we distinguish open-weight from open-source LLMs (see \emph{Recommendations} above).
    \item \emph{Performance gap.} Open models often lag behind the most advanced proprietary ``frontier'' models in common benchmarks, making it difficult to demonstrate clear improvements when evaluating new methods using open LLMs alone.
    \item \emph{Hardware demands.} Deploying and experimenting with these models typically requires substantial hardware resources, in particular high-performance GPUs that may be beyond reach for many academic groups.
    \item \emph{Operational complexity.} Unlike APIs provided by proprietary vendors (e.g., the OpenAI API), installing, configuring, and fine-tuning open models can be technically demanding.
\end{itemize}

\guidelinesubsubsection{Study Types}

This guideline applies primarily to study types in which the researcher controls which LLM is used.
For \benchmarkingtasks, an open LLM \should be one of the models under evaluation, so that the reported scores can be independently re-run. Where this is not feasible, researchers \should justify the omission and, per \design, ensure the evaluation harness can be used with open models.
For controlled experiments under \llmusage, an open LLM \should be one of the models under test; if the experimental design requires capabilities that only a specific commercial model exhibits, researchers \should acknowledge this as a limitation.
When evaluating \newtools, researchers \should use an open LLM as a baseline whenever it is technically feasible; if integration proves too complex, they \should report the initial benchmarking results of open models.
For \annotators and \judges, researchers \should compare annotation or judgment quality from open vs.\ commercial models to assess the extent to which results depend on a specific proprietary model.
For \synthesis, researchers \should compare synthesis results from open and commercial models to evaluate robustness of the findings.
For \llmusage, using an open LLM as a baseline is often not feasible when the study observes participants using specific commercial tools; in such cases, investigators \should explicitly acknowledge its absence and discuss how this limitation might affect their conclusions.
For \subjects, using an open LLM as a baseline may similarly be impractical if the study design requires the capabilities of a specific model; researchers \should acknowledge this limitation when applicable.

\guidelinesubsubsection{Advice for Reviewers}

Reviewers should distinguish between LLM use that is central to the research (e.g., building LLM-driven SE tools, using LLMs for synthesis) and use that is tangential (e.g., generating recruiting materials). An open LLM baseline is expected only when LLM use is central; otherwise, its absence need not be justified.
When an open baseline is expected, reviewers should look for either the use of an open LLM or a convincing argument for why it is impractical. If authors claim openness, some justification for that characterization is appropriate. Where practical, reviewers should examine whether the replication package contains sufficiently detailed instructions.
Performance differences between open and proprietary models are beyond the study authors' control. Reviewers should focus on whether the open baseline serves the study's methodological purpose rather than on its absolute performance.

\guidelinesubsubsection{See Also}
\begin{itemize}[label=$\rightarrow$]
    \item \seemodelversion: Authors must report version, configuration, and parameters for open models, not just commercial ones.
    \item \seedesign: Self-hosting an open model adds hardware, hosting, and infrastructure to document.
    \item \seebenchmarksmetrics: Inter-model agreement metrics make open-vs-commercial comparisons interpretable.
    \item \seelimitationsmitigations: When using an open LLM as a baseline is impractical, authors must report this as a limitation.
\end{itemize}
\space%

\subsection{Use Human Validation for LLM Outputs}
\label{sec:use-human-validation-for-llm-outputs}

\vspace{0.5\baselineskip}
\begin{framed}
\summary: %
If assessing the quality of generated artifacts is important and no reference datasets or suitable comparison metrics exist, researchers \should use human validation for LLM outputs. If they do, they \must define the measured construct (e.g., usability, maintainability) and describe the measurement instrument in the \paper. Researchers \should consider human validation early in the study design, and not as an afterthought. They \should build on established reference models for human-LLM comparison. When developing or adapting measurement instruments, researchers \must share them. When LLMs replace, rather than augment, humans in research tasks such as annotating software artifacts, coding interview transcripts, or simulating study participants, researchers \must explain whether and how the replacement is justified and \should ground it with inter-model and model-to-human agreement. For qualitative coding of interpretive data, they \should justify why an LLM is methodologically appropriate. When aggregating LLM judgments, methods and rationale \should be reported and inter-rater agreement \should be assessed. Confounding factors \should be controlled for; where applicable, researchers may perform a power analysis to estimate the required sample size. For value-laden or culturally contingent constructs, researchers \should describe rater demographics beyond expertise and discuss potential demographic biases. When evaluating agentic tools, user feedback on the agent's proposed actions \should be assessed and acceptance statistics reported.
\end{framed}

\guidelinesubsubsection{Rationale}

Even with well-justified benchmarks and metrics (\benchmarksmetrics), automated measurement captures only the constructs the benchmark was designed to measure.
Subjective constructs and constructs that no current benchmark covers require human assessment.

\guidelinesubsubsection{Recommendations}

When LLMs automate research or software development tasks previously performed by humans, the LLM's performance needs to be assessed.
Where no benchmark adequately operationalizes the target construct, researchers \should validate LLM outputs against human judgment.

\guidelineparagraph{Study Design Considerations}
Studies that include human participants need additional considerations, including a recruitment strategy, annotation guidelines, training sessions, or ethical approvals.
Therefore, researchers \should consider human validation early in the study design, not as an afterthought.
Authors \must clearly define in the \paper the constructs that the human and LLM annotators evaluate~\cite{DBLP:conf/ease/RalphT18}.
When designing custom instruments to assess LLM output (e.g., questionnaires, scales), researchers \must share these instruments.

\guidelineparagraph{Replacing Human Judgment}
LLMs may be used to replace, rather than augment, humans in research tasks such as annotating software artifacts, coding interview transcripts, or simulating participants in user studies. In such cases, researchers \must explain whether and how the replacement is justified and \should follow systematic approaches to support this judgment, e.g., showing that a jury of three LLMs exhibits inter-model agreement at the same threshold researchers would require for human-to-human agreement, such as Krippendorff's $\alpha>0.8$. Such agreement is necessary but not sufficient. High inter-model agreement can reflect shared model biases rather than valid annotation, so researchers \should additionally validate a sample against human experts before treating LLM outputs as a substitute for human work~\cite{DBLP:conf/msr/AhmedDTP25, Krippendorff2018}. For qualitative coding of interpretive data such as open-ended developer interview responses, agreement metrics do not show that an LLM can perform the interpretive work that the task requires. Researchers \should justify why the LLM is methodologically appropriate (see \llmsforresearcher).

\guidelineparagraph{Subjective Judgment and Agreement}
When the judgment is subjective (i.e., depends on the judge's values or theories), the same artifacts \should be judged independently by both the LLM and a panel of human experts, and the LLM judgments \should then be compared against an aggregated human reference. Researchers \should clearly describe their aggregation method and reasoning.
Researchers \should use established reference models to compare humans with LLMs.
For example, \citeauthor{DBLP:conf/icse/SchneiderFW25} outline design considerations for studies comparing LLMs with humans~\cite{DBLP:conf/icse/SchneiderFW25}.

One systematic approach to building the human reference is to randomly order the objects requiring judgment and put them in groups small enough to judge in one to three hours. Two or three human experts review the first group together, simultaneously judging, discussing, and creating a set of decision rules documenting their reasoning. Then, the experts iterate among rounds of:
\begin{enumerate}
    \item independently rating one group of objects,
    \item calculating inter-rater agreement (IRA) or reliability (IRR),
    \item meeting to discuss disagreements and reach consensus,
    \item updating the decision rules so the disagreement will not recur.
\end{enumerate}

The goal is to reach a target IRA or IRR (e.g., Krippendorff's $\alpha>0.8$), indicating sufficient decision rules. Once this target is reached and sustained for two to three groups, it is permissible to continue with a single rater.
Researchers \should report measures of IRA or IRR, preferably broken down by round, and \should apply thresholds consistent with established standards. \citeauthor{Krippendorff2018} recommends discarding data with $\alpha<0.667$, considers $0.667 \leq \alpha < 0.8$ sufficient only for tentative conclusions, and requires $\alpha \geq 0.8$ for reliable data~\cite{Krippendorff2018}.

Confounding factors \should be discussed and, where feasible, controlled for (e.g., by categorizing participants according to their level of experience or expertise).
For value-laden or culturally contingent constructs (e.g., judging code-style appropriateness or comment helpfulness), researchers \should describe rater demographics beyond expertise (e.g., geographic, linguistic, or professional background) and discuss potential demographic biases in rater recruitment and instructions~\cite{bean2025measuring}.
Where applicable, researchers may perform a power analysis~\cite{Cohen1992,DBLP:journals/infsof/DybaKS06} to estimate the required sample size, ensuring sufficient statistical power in their experimental design.
Although established sample-size guidance for LLM-human comparisons is limited, related fields commonly use 100 comparisons without further justification~\cite{DBLP:journals/npjdm/TamSKSPMOWVFMCSPW24}.

\guidelineparagraph{Agentic Tools}
Agentic human-in-the-loop interaction is a built-in human validation, with larger degrees of freedom than traditional experiments that use LLMs directly.
Besides generating and modifying content, agentic systems such as \emph{Claude Code} (see \design) can autonomously call command-line tools or pull in additional information from MCP servers.
When evaluating agentic tools, researchers \should assess the feedback that users provided on the agent's proposed actions (e.g., file edits, command executions), report statistics on how frequently they accepted the proposals, and how they modified them.

\guidelinesubsubsection{Examples}

\citeauthor{DBLP:conf/msr/AhmedDTP25} proposed a systematic method for deciding whether LLMs can replace human annotators on a given task, using inter-model agreement as an initial screening criterion (with a threshold of $\alpha>0.5$) and model confidence for sample-level decisions~\cite{DBLP:conf/msr/AhmedDTP25}. However, their threshold is well below the levels generally considered acceptable for inter-rater agreement.
Notably, while three LLMs exhibited higher inter-model agreement than human annotators on some tasks, human-model agreement remained low on others.
This illustrates that high inter-model reliability does not guarantee alignment with human judgment, reinforcing the need for human validation before assuming that LLM annotations are valid.
Moreover, a high inter-model agreement could merely indicate that the models share systematic biases and hence reliably agree on the wrong answer.

\citeauthor{DBLP:journals/corr/abs-2501-19297} evaluated ChatGPT's capability to generate requirements documents by comparing an LLM-generated and a human-generated document based on the same business use case~\cite{DBLP:journals/corr/abs-2501-19297}. Domain experts reviewed both documents and attempted to distinguish their origin.
For agentic tools, \citeauthor{DBLP:conf/icse-seip/TakerngsaksiriPTTZJLCCW25} reported acceptance and modification rates from real users for \emph{HULA}, an agentic plan-and-code system deployed in Atlassian JIRA~\cite{DBLP:conf/icse-seip/TakerngsaksiriPTTZJLCCW25}.

\guidelinesubsubsection{Benefits}

Validation against human judgments builds confidence in the LLM's accuracy and validity~\cite{khraisha2024canlargelanguagemodelshumans}. Reporting IRA or IRR supports this validation but does not stand in for it, since even high agreement may reflect shared biases rather than valid judgment. When human reviewers disagree with the LLM, the disagreement points to concrete improvements in the prompts, the context provided to the LLM, or the operational definition of the construct.

\guidelinesubsubsection{Challenges}

Assessing LLM performance with respect to a construct such as code maintainability, answer correctness, or annotation reliability is challenging because ensuring that a construct is defined well and operationalized using an appropriate measurement model requires a deep understanding of (1) the construct, (2) construct validity in general, and (3) instrumentation~\cite{DBLP:journals/tse/SjobergB23,DBLP:books/sp/24/RalphKAA24}. Comparing an LLM to human judges is typically slower and more expensive than machine-generated measures. More fundamentally, neither human judgment nor machine-generated measures provides an objective ground truth against which LLM accuracy can be firmly determined. Human preference ratings under-represent properties such as factuality and faithfulness, and assertively phrased outputs receive systematically fewer flagged errors from crowdworkers~\cite{DBLP:conf/iclr/HoskingBB24}. Human validation is worth its added cost only when automated measures alone fail to capture the constructs the study evaluates.

Human judgments exhibit variability due to differences in experience, expertise, interpretations, and personal biases~\cite{DBLP:journals/pacmhci/McDonaldSF19}. In pairwise judgment of LLM outputs, both human and LLM judges shift their preferences toward answers carrying fake references or richly formatted content, regardless of correctness~\cite{DBLP:conf/emnlp/ChenCLJW24}. When diverse humans rate items reliably given clear decision rules, we assume that reliability implies validity, but it does not. Measuring reliability is \textit{much} easier than measuring validity, and often the best researchers can do is argue conceptually for why their judges, decision rules, and constructs \textit{should} produce valid ratings.

\guidelinesubsubsection{Study Types}

This guideline applies to all study types, although the need for human validation varies. \emph{Replacing Human Judgment} introduces a conditional \must that applies particularly to \annotators, \judges, \synthesis, and \subjects, where the LLM substitutes for human input.
For \annotators, researchers \should validate LLM-generated annotations against human annotators to assess labeling quality and identify systematic biases.
When using \judges, researchers \should co-create initial rating criteria with humans and validate a sample of LLM judgments against human expert assessments.
For \synthesis, researchers \should employ human oversight to verify that qualitative interpretations and synthesized outputs faithfully represent the underlying data.
For \subjects, researchers \should validate simulated responses against real human data to assess the fidelity of the simulation.
For \llmusage, researchers \should carefully reflect on the validity of their evaluation criteria and validate subjective assessments with human experts.
For \newtools whose output is supposed to match human expectations, researchers \should validate the LLM output against human judgment.
For \benchmarkingtasks, there is less need for human validation when using extensively validated and widely-used benchmarks, but researchers \should employ human validation when creating or adapting new benchmarks.

\guidelinesubsubsection{Advice for Reviewers}

Human validation may be the most challenging of our guidelines to assess because it often requires evaluating conceptual arguments. If LLM output is validated only by comparison with other LLMs, reviewers should look for \textit{quantitative empirical evidence} that such comparison is reliable \textit{and} valid. High inter-model agreement alone is insufficient, as reliability does not imply validity. Similarly, reviewers should expect evidence that any employed benchmarks are reliable and valid. Absent such evidence, human validation is warranted.
A single human judge is appropriate only when judgments depend on widely accepted theories and involve limited value conflict (e.g., tagging method names containing abbreviations). For multiple judges, reviewers should expect IRA/IRR improvement techniques as described in the recommendations above (experienced raters, organized rounds, consensus meetings, updated decision rules). Low IRA or IRR (e.g., Krippendorff's $\alpha<0.8$) without these techniques is a concern. Conversely, if authors have followed best practices and still obtained mediocre results (e.g., $0.66<\alpha<0.8$), this should be noted as a limitation.
Beyond reliability, reviewers should expect authors to explain conceptually why their human judgments should be valid, considering construct definitions, decision rules, and judge expertise.
As with other guidelines, missing information is typically a revision request. Absent judge instructions, instruments, decision rules, or construct definitions may prevent assessment of rigor and validity, while missing recruitment details are less critical. Clarification requests about construct definitions are routine and should not alone warrant rejection.

\guidelinesubsubsection{See Also}
\begin{itemize}[label=$\rightarrow$]
    \item \seebenchmarksmetrics: Human evaluation complements automated metrics where benchmarks cannot sufficiently capture the target construct.
    \item \seemodelversion: Human-validation findings characterize only the specific model and configuration that produced the outputs.
    \item \seedesign: Prompt and architecture choices produce the outputs that humans then validate.
    \item \seetraces: Stored interaction logs and runtime traces are the artifacts human reviewers examine.
    \item \seelimitationsmitigations: Rater demographics, threshold choices, and the validity of human judgments are limitations to discuss.
\end{itemize}

\subsection{Report Limitations and Mitigations}
\label{sec:report-limitations-and-mitigations}

\vspace{0.5\baselineskip}
\begin{framed}
\summary: %
Researchers \must transparently report study limitations, including the impact of non-determinism and generalizability constraints. The \paper~\must specify whether generalization across LLMs or across time was assessed, and discuss model and version differences. Authors \must discuss potential data leakage effects and their impact on results, including the risk of evaluation data entering model improvement pipelines, and \must describe how the quality of subjective results was ensured. For studies involving sensitive data, they \must discuss data governance mechanisms. They \should justify LLM usage in light of its resource demands. Mitigation strategies such as replication packages, human validation, longitudinal re-runs, triangulation, and sensitivity analysis \should be employed and reported where applicable. Where full data sharing is not possible, a subset of the validation data \should be included to enable partial replication.
\end{framed}

\guidelinesubsubsection{Rationale}

When using LLMs for empirical studies in SE, researchers face unique challenges and potential limitations that can influence the validity, reliability, and reproducibility of their findings~\cite{DBLP:conf/icse/SallouDP24}.
Researchers must openly discuss these limitations and explain how their impact was mitigated.
These limitations are relative to current LLM capabilities and tool architectures; speculating about future improvements is beyond the scope of a paper's limitation section.
Nevertheless, risk management and threat mitigation \should be planned during study design, not as an afterthought.

\guidelinesubsubsection{Recommendations}
Researchers \must clearly present the limitations of their work without defensiveness or obfuscation. These limitations may concern external, internal, and construct validity; reliability and reproducibility; and ethical, regulatory, and environmental concerns.

We follow the standard SE convention of organizing limitations by external, internal, and construct validity, plus reliability~\cite{DBLP:journals/ese/RunesonH09,ralph2021empiricalstandardssoftwareengineering}, extended below with ethical, regulatory, and environmental concerns specific to LLM research. For studies that adopt qualitative analytic methods (the use of LLMs in reflexive qualitative analysis is itself contested, see \annotators), researchers \should use the trustworthiness criteria of credibility, transferability, dependability, and confirmability instead~\cite{guba1981criteria}. When deterministic reproducibility is structurally unattainable (e.g., SaaS-based models with opaque versioning), researchers \should adopt the same trustworthiness criteria to substantiate dependability and confirmability of findings.

\guidelineparagraph{External Validity}
The primary threats to external validity are:
\begin{itemize}
    \item \emph{Cross-model transfer limitations}: Results obtained with one LLM or family of LLMs may not generalize to others due to differences in training data, architecture, and post-training procedures (e.g., fine-tuning and reinforcement learning from human feedback); see \openllm for using an open LLM as a comparison baseline.
    \item \emph{Configuration sensitivity}: Results may not generalize beyond the specific configuration(s) tested (e.g., decoding parameters, system prompt, or context settings); see \design for an overview.
    \item \emph{Tool-architecture specificity}: Tools built around vendor-specific APIs or features (e.g., function calling, structured output, or context-window size) may not transfer to other models without substantial re-engineering (see \design).
    \item \emph{Limited domain coverage}: Studies often focus on a narrow set of programming languages, task types, or application domains, limiting generalizability to other SE contexts.
    \item \emph{Limited participant and rater diversity}: Study participants (e.g., developers) and human raters validating LLM outputs may not represent the broader population in terms of expertise, geographic location, or cultural background (see \humanvalidation for guidance on rater diversity in value-laden constructs).
    \item \emph{Research-to-practice gap}: Developers using the same tools outside researcher-supervised conditions may obtain results that differ from those reported in the study.
    \item \emph{Cross-time instability}: Performance of proprietary models can change over time, leading to non-generalizable study outcomes~\cite{DBLP:journals/corr/abs-2307-09009, doi:10.1148/radiol.232411} (see \modelversion for version and fingerprint reporting that enables tracking such drift).
\end{itemize}

Generalizability is particularly critical for proprietary and non-deterministic systems whose behavior is subject to drift (i.e., silent changes in model output over time). Researchers \must discuss the limitations and mitigations of external validity.
Mitigations include \emph{triangulation} across multiple models (e.g., proprietary and open), independent datasets, and complementary metrics; \emph{sensitivity analysis} that varies LLM configurations, prompts, architecture decisions, datasets, and where applicable participant backgrounds; and performing \emph{longitudinal re-runs} (see \emph{Reliability \& Reproducibility} below), which also helps detect cross-time instability.

\guidelineparagraph{Internal Validity}
The primary threats to internal validity are:
\begin{itemize}
	\item \emph{Data leakage and contamination}: Inter-dataset duplication can produce training-evaluation overlap, yielding overly optimistic results (see \benchmarksmetrics).
	\item \emph{Evaluation data entering model-improvement pipelines}: Evaluation samples can unintentionally feed retraining or fine-tuning, especially in longitudinal studies involving LLMs.
	\item \emph{Incomplete architecture, prompt, or pipeline reporting}: Undisclosed components introduce hidden confounders (see \design).
\end{itemize}

As transparency on training data is limited for LLMs, researchers \must discuss potential data leakage effects and their impact on results. Concrete mitigations for contamination (e.g., post-cutoff benchmark construction, held-out subsets, canary strings) are discussed in \benchmarksmetrics.

\guidelineparagraph{Construct Validity}
The primary threats to construct validity are:
\begin{itemize}
	\item \emph{Metric-construct mismatch}: Traditional metrics such as BLEU or ROUGE may miss SE-specific aspects such as functional correctness or behavioral equivalence (see \benchmarksmetrics).
	\item \emph{Construct under-specification}: If a construct lacks an operational definition, neither automated metrics nor human raters can apply it consistently.
	\item \emph{Reliability without validity}: High inter-rater or inter-model agreement does not imply that the measurement captures the intended construct; a reliable LLM can be reliably inaccurate (see \humanvalidation).
	\item \emph{Benchmark scope limitations}: Benchmarks commonly ignore runtime behaviors, security implications, readability, testability, and maintainability, yielding results that may not transfer to realistic development settings.
	\item \emph{Capability confounding}: Benchmark performance can blend the target capability with unrelated capabilities such as output format compliance, instruction following, or prompt format sensitivity, inflating apparent scores (see \benchmarksmetrics).
	\item \emph{Over-reliance on benchmark-specific metrics}: Optimizing for a single benchmark may produce dataset-specific shortcuts that pass the benchmark without exhibiting the capability it was designed to test, overstating real-world utility.
	\item \emph{Prompt sensitivity}: Small changes in instructions, formatting, or in-context examples can substantially shift what the LLM appears to measure, making the operationalized construct unstable across prompt variants (see \design).
	\item \emph{Judge biases}: LLM and human judges exhibit systematic biases such as position bias, verbosity bias, and preferences for richly formatted or citation-rich outputs regardless of correctness (see \humanvalidation).
\end{itemize}

If constructs are based on subjective interpretations, purely automated metrics are insufficient. Researchers \must discuss how they ensured quality of subjective results, similarly to qualitative research. The primary mitigation is \emph{human validation} of subjective constructs following quality criteria known from qualitative research (see \humanvalidation).

\guidelineparagraph{Reliability \& Reproducibility}
The primary threats to reliability and reproducibility are:
\begin{itemize}
    \item \emph{Non-deterministic outputs}: Identical prompts and configurations can yield different outputs across runs due to factors such as floating-point arithmetic, batching, and stochastic decoding strategies.
    \item \emph{Infrastructure dependence}: Results may vary depending on the hardware, software stack, and hosting environment used; vendor-imposed quotas, throttling, or pricing changes can further prevent re-running experiments at the original scale, making exact replication challenging across different infrastructure setups.
    \item \emph{Resource inequality}: LLM research is resource-intensive and remains predominantly in the domain of private companies or well-funded research institutions~\cite{schwartz2020green, ahmed2023industry}, excluding researchers from under-resourced institutions.
\end{itemize}

Researchers \must discuss the measures taken to increase reliability and reproducibility.
However, non-deterministic reproducibility is not inherently disqualifying. The trustworthiness criteria introduced above apply particularly to SaaS-based LLM research, where providers frequently deprecate model versions without guaranteeing stable behavior.
Mitigations include providing \emph{replication packages} that cover prompt and architecture specifications, model outputs, and representative examples for partial replicability (ideally accompanied by an implementation using an open model for long-term stability), and performing \emph{longitudinal re-runs} with statistical analyses. When deterministic reproduction is structurally impossible, researchers \should consider \emph{methodological trustworthiness measures} such as triangulation, reflexivity, audit trails, and peer debriefing as complementary measures.

\guidelineparagraph{Ethical \& Regulatory Boundaries}
The primary concerns for ethical and regulatory matters are:
\begin{itemize}
    \item \emph{Use of sensitive or proprietary data}: Studies involving proprietary code, confidential business data, or personally identifiable information may face restrictions on data sharing that limit reproducibility.
    \item \emph{Jurisdictional obligations}: Data protection regulations such as GDPR or CCPA and institutional policies may impose constraints on data collection, processing, and sharing in LLM-based studies.
    \item \emph{Implicit model bias}: Especially for qualitative research, LLMs might \enq{reinforce dominant paradigms and biases} and \enq{identify, replicate and reinforce dominant language and patterns}~\cite{jowsey2025reject} (see \humanvalidation).
\end{itemize}

Studies involving sensitive data \must discuss data governance mechanisms tailored toward LLM environments, compliant with applicable jurisdictional obligations. If applicable, researchers \should discuss how model biases potentially impact the study outcomes and how those biases were evaluated.

\guidelineparagraph{Environmental \& Sustainability Constraints}
The primary environmental and sustainability concerns are:
\begin{itemize}
    \item \emph{Energy consumption}: With growing model size, the environmental impact of experiments with LLMs increases, and the substantial energy costs of LLM experiments warrant consideration in study design~\cite{DBLP:conf/acl/StrubellGM19}.
    \item \emph{Trade-off between repetition and sustainability}: Repeating experiments increases reliability but also energy consumption, requiring trade-offs during study design.
\end{itemize}

Researchers \should justify the LLM's resource consumption against the benefits over traditional approaches.
Mitigations include \emph{energy preservation} and \emph{cost accounting}. \emph{Energy preservation} involves selecting smaller or newer, less resource-intensive models and applying techniques such as input/output token reduction, model pruning, quantization, or knowledge distillation~\cite{mitu2024hidden} where feasible. Carbon footprint estimation is desirable, but still difficult. \emph{Cost accounting} tracks resource consumption by reporting tokens, service costs, or hardware specifications.

\guidelinesubsubsection{Examples}

\citeauthor{DBLP:conf/icse/SallouDP24} catalog three categories of LLM-specific threats to validity (i.e., closed-source models, implicit data leakage, and reproducibility) and pair each with concrete mitigation strategies (e.g., versioned model archives, metamorphic test data, multiple replication runs with variability metrics, and detailed execution metadata)~\cite{DBLP:conf/icse/SallouDP24}.
\citeauthor{DBLP:conf/icse/Du0WWL0FS0L24} pair each threat in their \emph{ClassEval} evaluation with a concrete mitigation: manually constructing the benchmark with multiple annotators to limit data leakage, piloting prompts on held-out tasks to control for prompt sensitivity, and reporting greedy-decoding results to control for non-determinism~\cite{DBLP:conf/icse/Du0WWL0FS0L24}.

\guidelinesubsubsection{Benefits}

Transparent reporting of limitations and mitigations helps readers calibrate confidence in the findings, makes explicit which threats were addressed and which remain open, and documents design decisions that other authors can borrow or refine. It also keeps a paper's claims proportionate to its evidence.

\guidelinesubsubsection{Challenges}

Identifying limitations one is not already aware of is the hardest part of writing a threats section, particularly for methodological threats outside the team's primary expertise. Publication and reviewing norms can pressure authors to downplay weaknesses, while page limits make exhaustive treatment impractical. Threat lists that recite generic LLM-research issues (e.g., model bias, non-determinism, or contamination) without showing how each one applies to specific design choices in this study leave reviewers unable to tell which risks actually applied.

The threats to validity framework itself is contested within SE. \citeauthor{DBLP:journals/infsof/VerdecchiaELRS23} argue that threats sections too often read as \enq{laundry-lists}~\cite{DBLP:journals/infsof/VerdecchiaELRS23}, \citeauthor{DBLP:conf/esem/LagoRSV24} corroborated this empirically across a decade of ICSE Distinguished Paper Award winners~\cite{DBLP:conf/esem/LagoRSV24}, and \citeauthor{DBLP:journals/tosem/RobillardAEGLNNSSS24} argue for refocusing the discussion on study design trade-offs rather than the standard validity categories~\cite{DBLP:journals/tosem/RobillardAEGLNNSSS24}.

\guidelinesubsubsection{Study Types}

Researchers \must follow this guideline for all study types. Transparently reporting limitations and mitigations is a universal requirement, but specific concerns vary by study type.
For \annotators, researchers \must discuss potential biases in label assignment, label reliability limitations, and sensitivity of annotations to prompt wording and model choice.
For \judges, researchers \must address measurement validity concerns, known biases such as position bias or verbosity bias, and the extent to which LLM judgments align with human expert assessments.
For \synthesis, researchers \must discuss the risk of contextual misinterpretation, potential loss of nuance in summarized or aggregated outputs, and reflexivity limitations inherent in using an LLM for qualitative interpretation.
For \subjects, researchers \must discuss the fundamental inability of LLMs to truly simulate human behavior, the risk of stereotype amplification, and the limited ecological validity of simulated responses.
For \llmusage, researchers \must discuss generalizability constraints across different tools and user populations, and acknowledge how observed usage patterns may not transfer to other contexts.
For \newtools, researchers \must discuss replicability constraints arising from dependencies on commercial models, the impact of model updates on tool behavior, and limitations of the evaluation setup.
For \benchmarkingtasks, researchers \must discuss potential data contamination, benchmark scope limitations, capability confounding, and the extent to which benchmark performance generalizes to real-world tasks.

\guidelinesubsubsection{Advice for Reviewers}

Reviewers should verify that the limitation section is comprehensive and appropriate for the specific study type, checking that:
(1) limitations address the specific threats relevant to the study type (e.g., label reliability for annotation studies, simulation fidelity for studies using LLMs as subjects);
(2) mitigations are concrete and correspond to identified limitations rather than being generic statements;
(3) the impact of LLM non-determinism on findings is discussed;
(4) generalizability constraints, across models, configurations, time periods, and populations, are acknowledged.
When important limitations are missing, reviewers should request they be added. The absence of a limitation section, or one that is formulaic or insufficiently specific, is a more serious concern than any individual missing limitation and may warrant a major revision.

\guidelinesubsubsection{See Also}
\begin{itemize}[label=$\rightarrow$]
    \item \seemodelversion: Authors can re-run with different models or configurations to check whether the results depend on those specific choices.
    \item \seedesign: Triangulation across architectures and prompts is one mitigation strategy.
    \item \seetraces: Stored session traces serve as a baseline against which authors can monitor LLM behavior drift over time.
    \item \seebenchmarksmetrics: Benchmark and metric choices are one source of construct-validity threats authors must discuss.
    \item \seeopenllm: An open LLM baseline mitigates cross-model transfer concerns by providing an independently reproducible comparison.
    \item \seehumanvalidation: When automated metrics cannot validly measure a construct, human validation is an alternative.
\end{itemize}

\section{Conclusion}

LLM non-determinism, opaque training data, and rapidly evolving models
threaten the reproducibility and replicability of empirical SE studies.
This paper presents a taxonomy of seven study types that organizes how
LLMs are used in SE research, together with eight guidelines for
designing and reporting such studies.
Each guideline distinguishes requirements (\must) from recommendations
(\should) and is contextualized by the study types it
applies to.
Our guidelines recommend that researchers:
\begin{enumerate*}[label=(\arabic*)]
\item declare LLM usage and role;
\item report model versions, configurations, and customizations;
\item document the system and prompt design beyond the model;
\item report session traces, i.e., interaction logs and runtime traces;
\item use suitable baselines, benchmarks, and metrics;
\item include an open LLM as a baseline;
\item validate LLM outputs against human judgment; and
\item articulate limitations and mitigations.
\end{enumerate*}
An applicability matrix maps guidelines to study types, and a reporting
checklist (Appendix~\ref{sec:checklist}) supports authors and reviewers.
The study types and guidelines were derived collaboratively by
22~co-authors, following a process similar to that of guidelines in
other disciplines (see, e.g.,~\cite{Begg1996}).

Following the quality dimensions of the \emph{SIGSOFT Empirical
Standards}~\cite{ralph2021empiricalstandardssoftwareengineering}, we
consider our guidelines
(1)~\emph{comprehensive}, as they cover seven study types that we
identified and were derived from 22 co-authors' collective expertise;
(2)~\emph{useful}, as they provide actionable recommendations with an
applicability matrix and a concrete reporting checklist
(Appendix~\ref{sec:checklist});
(3)~\emph{well-argued}, as each guideline pairs a dedicated
rationale with concrete examples from published SE studies; and
(4)~\emph{integrated with previous work}, as they build on and
complement established empirical SE standards while addressing
LLM-specific challenges.
Adopting these practices will not eliminate LLM non-determinism, but it
would make claims auditable and results easier to replicate and compare
across studies.
Because models and tools evolve, we maintain the study types and
guidelines as a living resource and invite the community to refine and
extend them (\href{https://llm-guidelines.org/}{llm-guidelines.org}).

\section*{Acknowledgments}

Our study types and guidelines are based on discussion sessions with researchers at the \emph{2024 International Software Engineering Research Network} (ISERN) meeting and at the \emph{2nd Copenhagen Symposium on Human-Centered Software Engineering AI} (supported by the Carlsberg Foundation with grant CF24-0693 and the Alfred P. Sloan Foundation with grant G-2024-22586 to Daniel Russo).
We thank Steffen Herbold and Alexander Serebrenik for their feedback on the guidelines.
LLM-based tools, including ChatGPT (GPT-5 and GPT-5.5) and Claude Code (Opus 4.6, 4.7, and 4.8; Fable 5), were used interactively for language editing and structural revision suggestions across all sections of this paper. All LLM-generated suggestions were reviewed and revised by the authors.

\begin{appendices}

\section{Applicability Matrix}
\nopagebreak
\providecommand{\rot}[1]{\rotatebox{60}{\textbf{#1}}}


\begin{center}
\captionof{table}{Applicability of guidelines to study types.
\iconM~=~the guideline's core recommendations \must be followed for this study type,
\iconS~=~\should be followed,
--~=~are not directly applicable.
Each guideline's study-type-specific guidance is detailed in the corresponding subsection.}
\label{tab:guideline-matrix}
\footnotesize
\setlength{\tabcolsep}{4pt}
\renewcommand{\arraystretch}{1.15}
\begin{tabular}{@{} l c c c c c c c @{}}
& \rot{\hyperref[sec:llms-as-annotators]{Annotators}} & \rot{\hyperref[sec:llms-as-judges]{Judges}} & \rot{\hyperref[sec:llms-for-synthesis]{Synthesis}} & \rot{\hyperref[sec:llms-as-subjects]{Subjects}} & \rot{\hyperref[sec:studying-llm-usage-in-software-engineering]{Usage}} & \rot{\hyperref[sec:llms-for-new-software-engineering-tools]{Tools}} & \rot{\hyperref[sec:benchmarking-llms-for-software-engineering-tasks]{Benchmarking}} \\
\midrule
\hyperref[sec:declare-llm-usage-and-role]{Declare Usage}\hspace{12pt}
  & \iconM 
  & \iconM 
  & \iconM 
  & \iconM 
  & \iconM 
  & \iconM 
  & \iconM 
  \\
\hyperref[sec:report-model-version-configuration-and-customizations]{Model Version}
  & \iconM 
  & \iconM 
  & \iconM 
  & \iconM 
  & \iconM 
  & \iconM 
  & \iconM 
  \\
\hyperref[sec:report-system-and-prompt-design]{Design}
  & \iconM 
  & \iconM 
  & \iconM 
  & \iconM 
  & \iconS 
  & \iconM 
  & \iconM 
  \\
\hyperref[sec:report-session-traces]{Traces}
  & \iconS 
  & \iconS 
  & \iconS 
  & \iconS 
  & \iconM 
  & \iconS 
  & \iconS 
  \\
\hyperref[sec:use-suitable-baselines-benchmarks-and-metrics]{Benchmarks \& Metrics}
  & \iconM 
  & \iconS 
  & \iconS 
  & \iconS 
  & \iconS 
  & \iconS 
  & \iconM 
  \\
\hyperref[sec:use-an-open-llm-as-a-baseline]{Open LLM}
  & \iconS 
  & \iconS 
  & \iconS 
  & --     
  & --     
  & \iconS 
  & \iconS 
  \\
\hyperref[sec:use-human-validation-for-llm-outputs]{Human Validation}
  & \iconS 
  & \iconS 
  & \iconS 
  & \iconS 
  & \iconS 
  & \iconS 
  & \iconS 
  \\
\hyperref[sec:report-limitations-and-mitigations]{Limitations}
  & \iconM 
  & \iconM 
  & \iconM 
  & \iconM 
  & \iconM 
  & \iconM 
  & \iconM 
  \\
\bottomrule
\end{tabular}
\end{center}
\space%

\section{Rationale and Recommendations}
\nopagebreak

\begin{center}
\captionof{table}{Rationale and key recommendations for each guideline. \iconM~=~\must, \iconS~=~\should.}
\label{tab:rationale-recommendations}
\footnotesize
\begin{tabularx}{\textwidth}{@{}>{\raggedright\arraybackslash}p{2.05cm}>{\hsize=.72\hsize\raggedright\arraybackslash}X>{\hsize=1.28\hsize\raggedright\arraybackslash}X@{}}
\toprule
\textbf{Guideline} & \textbf{Rationale} & \textbf{Core Recommendations} \\
\midrule
Declare Usage & Transparency enables informed assessment of scope and limitations. &
\iconM Declare which LLM, how it was used, and where in the research process. \\
\addlinespace
Model Version & Reproducibility requires precise identification of the system used in a study. &
\iconM Report exact version, date, configuration, and fine-tuning details.
\newline \iconS Report defaults, checksums, and quantization; motivate model choice; acknowledge commercial-model reproducibility limits. \\
\addlinespace
Design & Static artifacts determine the model's input on every call and must be documented in full. &
\iconM Describe system and agent architecture, infrastructure, prompts, agent configuration, tool catalog, and retrieval mechanisms.
\newline \iconS For LLM usage, describe the tool architecture to the extent accessible. \\
\addlinespace
Traces & Runtime traces make LLM and agent behavior verifiable despite non-determinism and tool opacity. &
\iconM For studies of LLM usage, share full interaction logs subject to privacy constraints.
\newline \iconS Otherwise, share interaction logs, runtime traces, and plans where feasible. \\
\addlinespace
Benchmarks \& Metrics & Meaningful evaluation requires reasoned valid measurement. &
\iconM Justify metric and benchmark choices; discuss their validity.
\newline \iconS Define the phenomenon and sampling strategy; isolate confounders; analyze errors; repeat experiments and report distributions. \\
\addlinespace
Open LLM & Reproducibility depends on access to the model under study. &
\iconS Include an open LLM as a baseline; ensure it is independently reproducible from supplementary material; for benchmarks, design the harness for use with open models. \\
\addlinespace
Human Validation & Automated metrics alone cannot ensure validity of subjective constructs. &
\iconM Define the measured construct and share custom measurement instruments.
\newline \iconM When LLMs replace humans in research tasks, explain whether and how the replacement is justified.
\newline \iconS When LLMs replace humans in research tasks, ground the replacement with inter-model and model-to-human agreement.
\newline \iconS Validate against human judgment with inter-rater reliability; describe rater demographics for value-laden constructs. \\
\addlinespace
Limitations & Honest acknowledgment of threats strengthens a study. &
\iconM Discuss threats to internal validity (data leakage), reliability (non-determinism), construct and external validity.
\newline \iconS Employ mitigations where possible. \\
\bottomrule
\end{tabularx}
\end{center}
\space%

\section{Reporting Checklist}
\label{sec:checklist}


The following checklist, inspired by CONSORT~\citep{Schulz2010}, summarizes actionable items from the guidelines.
The checklist is organized along typical paper sections.
Items marked \iconM are requirements (\must), and items marked \iconS are recommendations (\should).
Each item references its source guideline by short name.
Items annotated with \paper or \supplementarymaterial indicate where we expect the information to be reported.
Unmarked items may be reported either in the paper or as supplementary material.
Items prefixed with a bracketed tag apply only to studies with that characteristic (e.g., \condition{fine-tuning}, \condition{agents}).
Beyond these characteristic-tagged items, each guideline's \emph{Study Types} subsection lists study-type-specific recommendations where applicable.

\paragraph{\textbf{Introduction}}
\begin{itemize}[label={},leftmargin=*]
    \item \iconM Disclose any use of LLMs in the empirical study, specifying which LLM, how, and where it was used~(\refdeclareusage).
    \item \iconS Report in the \paper the purpose of using LLMs, the tasks they automate, and the expected benefits~(\refdeclareusage).
\end{itemize}

\paragraph{\textbf{Research Design and Methods}}\mbox{}\\
\noindent\textit{Model Selection and Configuration}
\begin{itemize}[label={},leftmargin=*]
    \item \iconM Report in the \paper the exact LLM model or tool version, the configuration, and the date of study execution~(\refmodelversion).
    \item \iconM \condition{fine-tuning} For fine-tuned models, describe in the \paper the fine-tuning goal, the dataset, and the procedure~(\refmodelversion).
    \item \iconS Report default parameters and explain model and version choices~(\refmodelversion).
    \item \iconS Report checksums and additional model properties where available; for commercial tools, openly acknowledge their reproducibility limits~(\refmodelversion).
    \item \iconS \condition{quantization} For quantized models, report the quantization level (e.g., 4-bit, 8-bit) and method (e.g., GPTQ or AWQ)~(\refmodelversion).
    \item \iconS \condition{fine-tuning} Compare base and fine-tuned models using suitable metrics and benchmarks; share fine-tuning data and weights as \supplementarymaterial (or justify in the \paper why they cannot be shared)~(\refmodelversion).
    \item \iconS \condition{commercial-models} Include an open-weight LLM, ideally an open-source LLM, as a baseline when using commercial models and report inter-model agreement~(\refopenllm).
\end{itemize}

\noindent\textit{System and Prompt Design}
\begin{itemize}[label={},leftmargin=*]
    \item \iconM Describe in the \paper the full architecture of LLM-based tools, including the role of the LLM, interactions with other components, and overall system behavior~(\refdesign).
    \item \iconM Specify whether zero-shot, one-shot, or few-shot prompting was used~(\refdesign).
    \item \iconM Specify prompt reuse across models and configurations~(\refdesign).
    \item \iconM If the LLM was used in a standalone setup, with prompts sent directly to a model and no pre- or post-processing, state this explicitly~(\refdesign).
    \item \iconM Publish all prompts or, when using templates, prompt templates with representative instances, including their structure, content, formatting, and variable components, as \supplementarymaterial; include representative examples in the \paper~(\refdesign).
    \item \iconM \condition{few-shot} For few-shot prompts, explain in the \paper how the examples were selected~(\refdesign).
    \item \iconM \condition{dynamic-prompts} For dynamically generated prompts, document the code or rules that assemble each prompt from runtime inputs~(\refdesign).
    \item \iconM \condition{restricted-sharing} For partially shared prompts, anonymize personal identifiers, replace proprietary code with placeholders, and clearly highlight modified sections~(\refdesign).
    \item \iconM \condition{context-files} Describe in the \paper any configuration mechanisms used (e.g., context files such as \texttt{CLAUDE.md} or \texttt{AGENTS.md}, skills, subagents, hooks, settings, rules)~(\refdesign).
    \item \iconM \condition{tool-use} Summarize in the \paper which tools were exposed to the model~(\refdesign).
    \item \iconM \condition{agents} If autonomous agents are used, specify agent roles, reasoning frameworks, and communication flows~(\refdesign).
    \item \iconM \condition{agents} For agentic systems that use external tools, distinguish the model's reasoning and outputs, its tool calls, and its interactions with users, the environment, or other agents~(\refdesign).
    \item \iconM \condition{context-augmentation} For retrieval-augmented generation (RAG) or related methods, describe how external data was retrieved, stored, and selected for inclusion in the model's context~(\refdesign).
    \item \iconM \condition{benchmarking} Describe the evaluation harness and infrastructure when it goes beyond bare model API calls (e.g., custom sandboxing, orchestration layers, or post-processing pipelines)~(\refdesign).
    \item \iconM \condition{benchmarking} For pre-defined prompts from existing benchmarks, specify the benchmark version, and disclose and justify any modifications to the prompts or evaluation setup~(\refdesign).
    \item \iconM \condition{latency-sensitive} For time-sensitive measurements, clarify whether local infrastructure or cloud services were used, including the specific hardware for local hosting (e.g., GPU model and VRAM) or the service tier for cloud APIs~(\refdesign).
    \item \iconS Justify substantive architectural choices where alternatives existed (e.g., agentic framework, tool catalog)~(\refdesign).
    \item \iconS Describe how the models were hosted and accessed~(\refdesign).
    \item \iconS Describe prompt development rationale and selection process~(\refdesign).
    \item \iconS Report prompt evolution and any LLM-suggested refinements~(\refdesign).
    \item \iconS Where legally possible, release the source code of the implementation under an open-source license~(\refdesign).
    \item \iconS \condition{few-shot} Include the concrete few-shot examples in the \supplementarymaterial~(\refdesign).
    \item \iconS \condition{participant-prompts} For user-authored prompts, describe how they were collected and analyzed~(\refdesign).
    \item \iconS \condition{long-prompts} Document input handling and token optimization strategies when prompts are long or complex~(\refdesign).
    \item \iconS \condition{restricted-sharing} If full prompt disclosure is not feasible, provide summaries or examples~(\refdesign).
    \item \iconS \condition{ensemble} For ensemble architectures, explain in the \paper the coordination logic between models~(\refdesign).
    \item \iconS \condition{context-augmentation} Report data preprocessing, versioning, and update frequency for stored data used for context augmentation~(\refdesign).
    \item \iconS \condition{context-files} Include all configuration artifacts (context files, skill folders, subagent files, hooks, settings, rules) as \supplementarymaterial~(\refdesign).
    \item \iconS \condition{tool-use} Include the tool catalog (names with purposes), tool schemas, and connected MCP servers as \supplementarymaterial~(\refdesign).
    \item \iconS \condition{benchmarking} Design the evaluation harness so it is usable with open models~(\refdesign).
\end{itemize}

\noindent\textit{Session Traces}
\begin{itemize}[label={},leftmargin=*]
    \item \iconM Where tool-native trace formats are used (e.g., Claude Code's session transcripts, LangGraph's state logs), describe the file format and report the tool version~(\reftraces).
    \item \iconM \condition{restricted-sharing} For traces containing sensitive information, anonymize personal identifiers, replace proprietary code with placeholders, and clearly highlight modified sections~(\reftraces).
    \item \iconS Use an open trace format with a documented schema where it fits (e.g., the OpenTelemetry GenAI semantic conventions or OpenInference)~(\reftraces).
    \item \iconS Include full interaction logs (prompts and responses) as \supplementarymaterial if privacy and confidentiality can be ensured~(\reftraces).
    \item \iconS \condition{agents} For agentic systems, include interaction logs covering human-in-the-loop exchanges with the agent (feedback, approvals, refinements) as \supplementarymaterial~(\reftraces).
    \item \iconS \condition{agents} For agentic systems, report the complete runtime trace as \supplementarymaterial, including for each entry the tool or artifact name, arguments, result, and ordering, and which configured artifacts (skills, context files, subagents) were activated~(\reftraces).
    \item \iconS \condition{agents} For agentic systems, report any plans the system exposes as \supplementarymaterial~(\reftraces).
\end{itemize}

\noindent\textit{Benchmarks and Metrics}
\begin{itemize}[label={},leftmargin=*]
    \item \iconM Justify in the \paper all benchmark and metric choices~(\refbenchmarks).
    \item \iconM Discuss in the \paper the reliability and validity, especially construct validity, of the selected benchmarks and metrics~(\refbenchmarks).
    \item \iconM Explain in the \paper why the selected metrics are suitable for the specific study; prior adoption in related work alone is not sufficient justification~(\refbenchmarks).
    \item \iconM \condition{latency-sensitive} Report latency when it can affect study outcomes (e.g., interactive user studies, latency comparisons)~(\refbenchmarks).
    \item \iconM \condition{new-benchmark} For new or updated benchmarks, disclose data sources and collection dates for each release~(\refbenchmarks).
    \item \iconS Provide an operational definition of the phenomenon the benchmark is intended to measure, including its scope and any sub-components~(\refbenchmarks).
    \item \iconS Summarize benchmark structure, task types, and limitations~(\refbenchmarks).
    \item \iconS Identify the capabilities a benchmark conflates with the target phenomenon, isolate the target where possible, and acknowledge remaining confounders as construct-validity threats~(\refbenchmarks).
    \item \iconS Perform an error analysis: categorize the failures observed and report their relative frequency; report failures that cluster on confounding capabilities as construct-validity threats~(\refbenchmarks).
    \item \iconS Describe and justify the sampling strategy used to select problems for inclusion in the benchmark~(\refbenchmarks).
    \item \iconS Justify the number of experiment repetitions, for example through a power analysis or by monitoring convergence of descriptive statistics~(\refbenchmarks).
    \item \iconS \condition{non-probability-sampling} For non-probability sampling (e.g., convenience), discuss the implications for the generalizability of conclusions~(\refbenchmarks).
    \item \iconS \condition{new-benchmark} For new or released benchmarks, adopt contamination-prevention mechanisms: held-out subset, canary strings, and pre-exposure investigation against common training corpora~(\refbenchmarks).
    \item \iconS \condition{multi-rater-scoring} For ratings that vary across raters or runs (human raters, LLM-as-judge), report the distribution of ratings per item rather than only aggregated point estimates~(\refbenchmarks).
\end{itemize}

\noindent\textit{Human Validation}
\begin{itemize}[label={},leftmargin=*]
    \item \iconM \condition{human-validation} If using human validation, define in the \paper the measured construct (e.g., usability, maintainability) and describe the measurement instrument~(\refhumanvalidation).
    \item \iconM \condition{human-validation} When developing or adapting measurement instruments, share them~(\refhumanvalidation).
    \item \iconM \condition{human-validation} When LLMs replace humans in research tasks, explain whether and how the replacement is justified~(\refhumanvalidation).
    \item \iconS Consider human validation early in the study design and build on established reference models for human-LLM comparison~(\refhumanvalidation).
    \item \iconS \condition{human-validation} When LLMs replace humans in research tasks, report the systematic approach used to justify the replacement, including inter-model and model-to-human agreement~(\refhumanvalidation).
    \item \iconS \condition{human-validation} Validate LLM judgments against human judgment, report aggregation methods, and assess human-LLM agreement~(\refhumanvalidation).
    \item \iconS \condition{human-validation} Discuss and, where feasible, control for confounding factors~(\refhumanvalidation).
    \item \iconS \condition{human-validation}\condition{subjective-constructs} For value-laden or culturally contingent constructs, describe rater demographics beyond expertise and discuss potential demographic biases~(\refhumanvalidation).
    \item \iconS \condition{agents} When evaluating agentic tools, assess the feedback users provided on the agent's proposed actions, and report how frequently proposals were accepted and how they were modified~(\refhumanvalidation).
\end{itemize}

\noindent\textit{Reproducibility, Ethics, and Resources}
\begin{itemize}[label={},leftmargin=*]
    \item \iconM \condition{restricted-sharing} For studies involving sensitive data, discuss data governance mechanisms compliant with applicable jurisdictional obligations~(\reflimitations).
    \item \iconS Justify LLM usage in light of its resource demands~(\reflimitations).
    \item \iconS Ensure the open-LLM baseline is independently reproducible from the \supplementarymaterial~(\refopenllm).
    \item \iconS \condition{restricted-sharing} Where full sharing of prompts, traces, or datasets is not feasible, share representative examples for partial replicability~(\reflimitations).
\end{itemize}

\paragraph{\textbf{Results}}
\begin{itemize}[label={},leftmargin=*]
    \item \iconS Repeat experiments due to the inherent non-determinism of LLMs and report the result distribution using descriptive statistics~(\refbenchmarks).
    \item \iconS Use traditional (non-LLM) baselines for comparison where possible~(\refbenchmarks).
    \item \iconS Report established metrics to make study results comparable; additional metrics may be reported where appropriate~(\refbenchmarks).
    \item \iconS \condition{comparing-models} If comparing models or tools, use appropriate inferential statistics (e.g., hypothesis tests, effect sizes) rather than relying solely on summary statistics~(\refbenchmarks).
\end{itemize}

\paragraph{\textbf{Limitations and Threats to Validity}}
\begin{itemize}[label={},leftmargin=*]
    \item \iconM Discuss potential data leakage effects and their impact on results, and describe how the quality of subjective results was ensured~(\reflimitations).
    \item \iconM Transparently report study limitations, including the impact of non-determinism and generalizability constraints~(\reflimitations).
    \item \iconM Specify whether generalization across LLMs or across time was assessed, and discuss model and version differences~(\reflimitations).
    \item \iconM \condition{restricted-sharing} Acknowledge non-disclosed confidential or proprietary components as reproducibility limitations~(\refdesign).
    \item \iconS Employ and report strategies to mitigate identified validity and reproducibility threats, such as replication packages, human validation, longitudinal re-runs, triangulation, and sensitivity analysis~(\reflimitations).
\end{itemize}
\space%

\end{appendices}
\space%

\section{Statements and Declarations}

\begin{description}
\item[\textit{Competing Interests:}] The authors declare no competing interests. Multiple co-authors are members of the EMSE Editorial Board, which is disclosed here for transparency.
\item[\textit{Authorship:}] Each author contributed or reviewed content and, most importantly, read the complete guidelines and confirmed that they stand behind all recommendations, not only their own contributions.
\item[\textit{Funding:}] No funding was received for writing this article. However, the \emph{2nd Copenhagen Symposium on Human-Centered Software Engineering AI}, which played a central role in forming the team of co-authors, was supported by the Carlsberg Foundation with grant CF24-0693 and the Alfred P.\ Sloan Foundation with grant G-2024-22586 to Daniel Russo.
\item[\textit{Ethical Approval:}] No ethics approval was required for writing this article.
\item[\textit{Data Availability:}] This article has no associated data. The guidelines are \href{https://llm-guidelines.org/}{hosted online} and \href{https://github.com/se-uhd/llm-guidelines-website}{maintained on GitHub}.
\end{description}

\bibliographystyle{spbasic}
\bibliography{literature}

\end{document}